\documentclass[preprint,10pt]{aastex}
\usepackage[utf8]{inputenc}
\usepackage{natbib}
\usepackage{booktabs}

\newcommand{\pname}{JS 183}   
\newcommand{\kepler}{{\it Kepler}}
\newcommand{\ktwo}{{\it K2}}
\newcommand{\teff}{\ensuremath{T_{\rm eff}}}
\newcommand{\msun}{\ensuremath{\,M_\Sun}}
\newcommand{\rsun}{\ensuremath{\,R_\Sun}}
\newcommand{\mj}{\ensuremath{\,M_{\rm J}}}
\newcommand{\rj}{\ensuremath{\,R_{\rm J}}}

\usepackage[normalem]{ulem}
\usepackage{color}
\newcommand{\kgsdel}[1]{\textcolor{red}{}}

\usepackage{pdflscape}

\begin{document}

\title{A Low-Mass Exoplanet Candidate Detected By \ktwo\ Transiting the Praesepe M Dwarf JS 183}
\author{
Joshua Pepper$^{1,2}$,
Ed Gillen$^{3}$, 
Hannu Parviainen$^{4}$, 
Lynne A. Hillenbrand$^{5}$, 
Ann Marie Cody$^{6}$, 
Suzanne Aigrain$^{3}$, 
John Stauffer$^{7}$, 
Frederick J. Vrba$^{8}$,
Trevor David$^{5,9}$,
Jorge Lillo-Box$^{10}$,
Keivan G. Stassun$^{2,11}$,
Kyle E. Conroy$^{2}$,
Benjamin J.\ S.\ Pope$^{3}$, 
David Barrado$^{12}$
}

\affil{$^1$ Department of Physics, Lehigh University, 16 Memorial Drive East, Bethlehem, PA 18015, USA}
\affil{$^2$ Department of Physics and Astronomy, Vanderbilt University, 6301 Stevenson Center, Nashville, TN 37235, USA}
\affil{$^3$ Astrophysics Group, Cavendish Laboratory, J.J. Thomson Avenue, Cambridge CB3 0HE, UK}
\affil{$^4$ Oxford Astrophysics, University of Oxford, Denys Wilkinson Building, Keble Rd, Oxford OX1 3RH, UK}
\affil{$^5$ Department of Astronomy, California Institute of Technology, 1200 E. California Blvd., MC 249-17, Pasadena, CA 91125, USA}
\affil{$^6$ NASA Ames Research Center, Mountain View, CA 94035, USA}
\affil{$^7$ Spitzer Science Center, California Institute of Technology, Pasadena, CA 91125, USA}
\affil{$^8$ U.S. Naval Observatory, Flagstaff Station, 10391 W. Naval Observatory Road, Flagstaff, AZ 86005-8521, USA }
\affil{$^9$ NSF Graduate Research Fellow}
\affil{$^{10}$ European Southern Observatory (ESO), Alonso de C\'ordova 3107, Vitacura, Casilla 19001, Santiago de Chile, Chile}
\affil{$^{11}$ Department of Physics, Fisk University, Nashville, TN 37208, USA}
\affil{$^{12}$ Depto. de AstrofM-CM--sica, Centro de AstrobiologM-CM--a (CSIC-INTA), ESAC campus 28691 Villanueva de la CaM-CM-1ada (Madrid), Spain}

\begin{abstract}
We report the discovery of a repeating photometric signal from a low-mass member of the Praesepe open cluster that we interpret as a Neptune-sized transiting planet.  The star is JS 183 (HSHJ 163, EPIC 211916756) with $\teff = 3325\pm100$ K, $M_{*} = 0.44\pm0.04$\msun, $R_{*} = 0.44\pm0.03$\rsun, and $\log{g_*} = 4.82\pm0.06$.  The planet has an orbital period of 10.134588 days and a radius of $R_{P}= 0.32\pm0.02$\rj.  Since the star is faint at $V=16.5$ and $J=13.3$, we are unable to obtain a measured radial-velocity orbit, but we can constrain the companion mass to below about 1.7\mj, and thus well below the planetary boundary.  JS 183b (since designated as K2-95b) is the second transiting planet found with \ktwo\
that resides in a several hundred Myr open cluster; both planets orbit mid-M dwarf stars and are approximately Neptune-sized.  With a well-determined stellar density from the planetary transit, and with an independently known metallicity from its cluster membership, \pname\ provides a particularly valuable test of stellar models at the fully convective boundary. We find that \pname\ is the lowest-density transit host known at the fully convective boundary, and that its very low density is consistent with current models of stars just above the fully convective boundary but in tension with the models just below the fully convective boundary.
\end{abstract}

\section{Introduction and Context}

Open clusters have been one of the preferred locations for exoplanet searches since the early days of exoplanet discovery.  Representing relatively compact, coeval populations of stars with similar composition, they serve as desirable testbeds for comparing planet frequencies and properties across various stellar parameters.  They have been especially selected for transit searches, since many open clusters are compact on the sky, requiring few separate telescope pointings.

Early transit searches for planets in clusters failed to detect planets \citep[e.g.][]{Mochejska:2005,vonBraun:2005,Burke:2006,Mochejska:2006,Aigrain:2007b,Mochejska:2008,Pepper:2008,Hartman:2009}.  It appears that the primary reason for this is that such searches were sensitive only to hot Jupiters, which are intrinsically rare \citep{Gould:2006,Wright:2012}, and that most open clusters do not have enough members for high probability detection of even one hot Jupiter \citep{Pepper:2005,Beatty:2008,Aigrain:2007a,vanSaders:2011}.  Radial-velocity (RV) searches suffer less from the low probability of fortuitous orbital inclination angles required for transit searches. Via an RV survey, \citet{Quinn:2012} discovered two Jupiter-mass planets in the open cluster Praesepe, with \citet{Malavolta:2016} finding an additional much higher mass planet in one of these two systems, and \citet{Quinn:2014} reported a Jupiter-mass planet in the Hyades cluster. Three additional Jupiter-mass RV planets were reported in M67 by \citet{Brucalassi:2014}.  Planets have also been found orbiting evolved members of open clusters \citep{Sato:2007,Lovis:2007}. 

The higher frequency of smaller planets compared to hot Jupiters \citep{Howard:2012,Petigura:2013} presents an opportunity to revisit open clusters as transit search targets.  Missions like Kepler \citep{Borucki:2010} have the photometric precision to detect small planets, and the repurposed Kepler mission \ktwo\ \citep{Howell:2014} covers a much larger fraction of the sky, allowing for multiple open clusters to be observed.  The first transiting planets orbiting stars in clusters were claimed by \citet{Meibom:2013}, who reported that two members of the $\sim$1-Gyr open cluster NGC 6811 in the Kepler field host sub-Neptune sized planet candidates associated with G-type main sequence stars. From \ktwo\ mission data, both \citet{Mann:2016hyades} and \citet{David:2016} reported the independent discovery of the Neptune-sized planet known as K2-25b orbiting the $\sim$600-800 Myr old Hyades cluster member Han87, with a late-M spectral type.  

There are benefits to detecting planets in young (sub-Gyr) clusters.  At early ages, the dynamical environments of the planetary systems might not be settled down.  Planet-planet interactions might still be occurring at that stage.  Once a statistically significant sample of extrasolar planets is found in clusters with various ages, we can explore the evolution of planetary systems at early times, with implications for planetary migration and dynamical stability.  That process might also provide insight into the actual locations in the disk where the planets originally formed.

M dwarfs are of specific interest as transiting planet hosts, given the larger transit signal afforded by smaller stars.  Also, it appears that M dwarfs are more likely than earlier-type stars to host planets \citep{Dressing:2013}, with implications for the frequency of planets in the habitable zones of M dwarfs \citep{Dressing:2015}.  According to the NASA Exoplanet Archive\footnote{\url{http://exoplanetarchive.ipac.caltech.edu/index.html}}, there are only 25 known M dwarfs (T$_{\rm eff} \le 3850$ K) with confirmed transiting planets.  Since M dwarf properties are notoriously difficult to pin down \citep[e.g., activity can inflate them, empirical relations for M dwarf radii suggest metallicity dependence, etc.;][]{Stassun:2012,Mann:2016}, this limits the precision on the transiting planet radii. Indeed, of the 25 previously known M dwarfs with transiting planets, only two (GJ 1132, GJ 1214) have measured parallaxes permitting a direct determination of the stellar radius and a spectroscopically determined metallicity to permit the most reliable stellar mass estimates from empirical relations, and because they are field stars, it is not possible to determine robust stellar ages. The expected arrival of parallaxes from the upcoming Gaia second data release promises to provide precise stellar and planetary parameters for many more transiting systems \citep[see, e.g.,][]{Stassun:2017}.

In this paper we report the discovery of a Neptune-sized planet transiting an M dwarf member of the Praesepe open cluster, observed by the \ktwo\ mission.  The star is alternately referred to as JS 183 \citep{JonesStauffer:1991}, HSHJ 163 \citep{Hambly:1995}, 2MASS J08372705+1858360 \citep{Cutri:2003,Skrutskie:2006}, and EPIC 211916756 \citep{Huber:2015}.  In the course of this work, this star was examined by other researchers and given the official \ktwo\ name K2-95b; however, since this star has been known as JS 183 in the literature for some time, we use the name JS 183 in this paper.  Since the host star is a member of a well-studied stellar population with known distance, metallicity, and age, it is possible to obtain precise stellar and therefore planet properties and to assign a precise age to the system. Section 2 discusses the properties of the host star.
Section \ref{sec:data} presents the archival and newly obtained photometric, spectroscopic, and imaging data that we use in this analysis. Section \ref{sec:results} presents the main results of this study, including a precise transit model, a detailed spectral energy distribution (SED) analysis providing precise stellar properties, an analysis of RV observations placing an upper limit on the planet mass, and the final adopted planet properties, showing the planet to be a Neptune-sized planetary mass object transiting an M3.5-type cluster star. We summarize our findings and conclusions in Section \ref{sec:summary}. 

\section{Praesepe Membership and Stellar Properties from the Literature}

\begin{deluxetable}{lcc}
\tablewidth{0pt}
\tabletypesize{\footnotesize}
\tablecolumns{3}
\tablecaption{Basic properties of \pname \label{tab:basic}}
\tablehead{
    \colhead{Property} & \colhead{Value} & \colhead{References} 
    }
\startdata
Names & JS 183, HSHJ 163, 2MASS J08372705+1858360, EPIC 211916756 & 1, 2, 3, 4 \\
R.A. (J2000.0) & 08:37:27.06 & 3 \\
Dec. (J2000.0) & +18:58:36.07 & 3 \\ 
Spectral Type & M3.5e & 5 \\ 
${\rm [Fe/H]}$ & 0.1--0.27 & 6 \\
V-mag & 17.08 & 7  \\
\enddata
\tablerefs{1: \citet{JonesStauffer:1991}, 2: \citet{Hambly:1995}, 3:  \citet{Cutri:2003,Skrutskie:2006}, 4: \citet{Huber:2015}, 5: \citet{Adams:2002}, 6: \citet{Pace:2008} 7: Rebull et al. 2016 (in prep)}
\end{deluxetable}

\pname\ is a modestly active M dwarf \citep[spectral type $\approx$M3.5e;][]{Adams:2002}, high-probability member of the Praesepe cluster. In this section we briefly summarize the evidence for cluster membership and chromospheric activity, as well as salient general properties of the Praesepe cluster that therefore apply to \pname. Table~\ref{tab:basic} summarizes the basic properties of \pname, including a digest of the various identifiers with which it has appeared in the literature.  The \ktwo\ name is EPIC 211916756.   

\subsection{Cluster Membership}

A number of focused studies and all-sky surveys have determined cluster membership probabilities for \pname\ based on proper motion and/or RVs. \citet{Kraus:2007} included \pname\ in their focused proper-motion study of the Praesepe cluster, finding a membership probability of 99.4\%.  \citet{Boudreault:2012} similarly combined UKIDSS proper motions and photometry to obtain a cluster membership probability of 92\%. 

\pname\ has also been included in all-sky proper-motion surveys, with the following proper motions ($\mu_\alpha$, $\mu_\delta$) [mas yr$^{-1}$] from USNO B1, PPMX, and URAT, respectively:
($-34$, $-14$), ($-39$, $-17$), ($-34$, $-17$). 
These agree well with the mean Praesepe cluster motion, which has been reported as ($-29$, $-7$), ($-34$, $-7.5$), and ($-36$, $-13$) from \citet{Adams:2002}, \citet{Boudreault:2012}, and \citet{Kraus:2007}, respectively. 
In summary, there is strong consensus in the literature that \pname\ is a high-probability member of the Praesepe cluster based on its proper motion and photometry.

\subsection{Cluster Age and Metallicity}

The accepted age of the Praesepe cluster is $\approx$600 Myr \citep[e.g.,][]{Adams:2002} although an older age of $\approx$800 Myr was recently advocated \citep{Brandt:2015}. 
Praesepe has been found to have a super-solar metallicity, with most values around [Fe/H] $\approx 0.1$ \citep[for a good summary see][]{Boesgaard:2013} but with some values as high as [Fe/H] = $0.27 \pm 0.10$ \citep[e.g.,][]{Pace:2008}.
In our analysis we make use of the metallicity to estimate the stellar radius expected from empirical M dwarf radius relations, and we use the cluster age to place the \pname\ planet system in the context of other systems of known age.

\subsection{Stellar Activity}
\label{sec:activity}

\citet{Barrado:1998} conducted a comparison of Praesepe and Hyades chromospheric/coronal activity. Their reported data for \pname\ (identified as HSHJ 163) from a December 1997 Keck/HIRES observation indicated a Balmer line equivalent width EW(H$\alpha$) = 0.00 \AA, and x-ray luminosity $\log L_X < 28.07$ erg s$^{-1}$. \citet{Adams:2002} used low-resolution spectroscopy to obtain a spectral type of M3.5e, indicating activity, with EW(H$\alpha$) = $-0.3$ \AA.  \citet{Kafka:2006} also report an  EW(H$\alpha$) = $-0.16$ \AA. In summary, all of these studies show H$\alpha$ just very weakly in emission or filled in, suggesting modest chromospheric activity.

\section{New Observations and Analysis\label{sec:data}} 

\subsection{\ktwo\ Observations and Light Curve Properties}

The photometric observations from \ktwo\ require considerable attention to the construction of proper apertures and removal of systematic noise (see, e.g. \citet{Vanderburg:2014,Foreman-Mackey:2015,Aigrain:2016}).  
Here we describe the extraction of the light curve of \pname, the removal of systematics, the identification of the transit signal, and the characterization of out-of-transit variability.

\subsubsection{Light curve preparation}

\begin{figure*}[ht]
    \centering
    \includegraphics[width=6in]{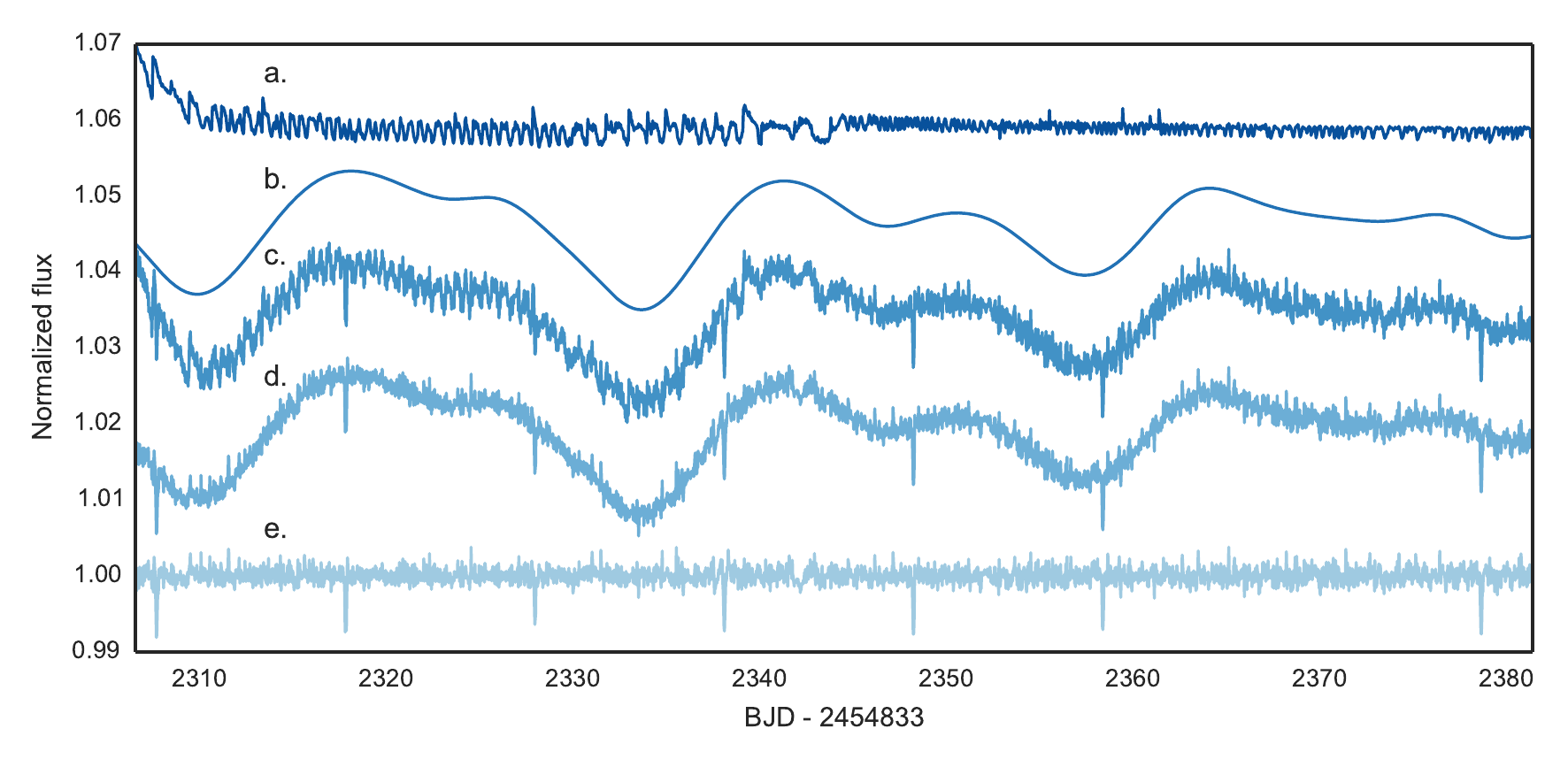}
    \caption{\ktwo\ light curve of \pname. From the top are the \textsc{K2SC} models of the pointing systematics (a), and of the stellar variability (b).  Then the original PDC light curve (c), the systematics corrected version (d), and the fully detrended version -- after removal of the stellar variability model (e).  Version (d) was used in the transit search and modelling described in \S \ref{sec:GP-transit}, while version (e) was used in the modelling described in \S \ref{sec:hannu}.  Note that there was a short gap in the original data at the time of the penultimate transit, which is why it appears to be missing.}
    \label{fig:K2_lc}
\end{figure*}

\pname\ was observed by \ktwo\ in its Campaign~5 field, from MJD 57139.13 to 57213.93, acquiring 3448 photometric observations at the standard 30-minute cadence. We downloaded the light curve file for \pname\ from the Mikulski Archive for Space Telescopes (MAST), which includes both Simple Aperture Photometry (SAP) fluxes, and a version of the light curve after application of the \emph{Kepler} Pre-search Data Conditioning (PDC) pipeline \citep{2010SPIE.7740E..1UT,2012PASP..124..985S,2012PASP..124.1000S}. 
This algorithm effectively models trends that are common to the ensemble of light curves, which was sufficient for the nominal \kepler\ mission but does not account for the diversity of pointing-related systematics present in \ktwo\ (see \citet{Vanderburg:2014} and \citet{vancleve:2016} for details). We therefore additionally used the \textsc{K2SC} Gaussian Process-based systematics correction algorithm of \citet{Aigrain:2016} to detrend both SAP and PDC light curves with respect to pointing variations. This algorithm  models the stellar variability simultaneously with the pointing-dependent systematics, and outputs a model for each component, so that either or both can be subtracted at will. 

\begin{figure}[ht]
    \centering
    \includegraphics[height=5.19in]{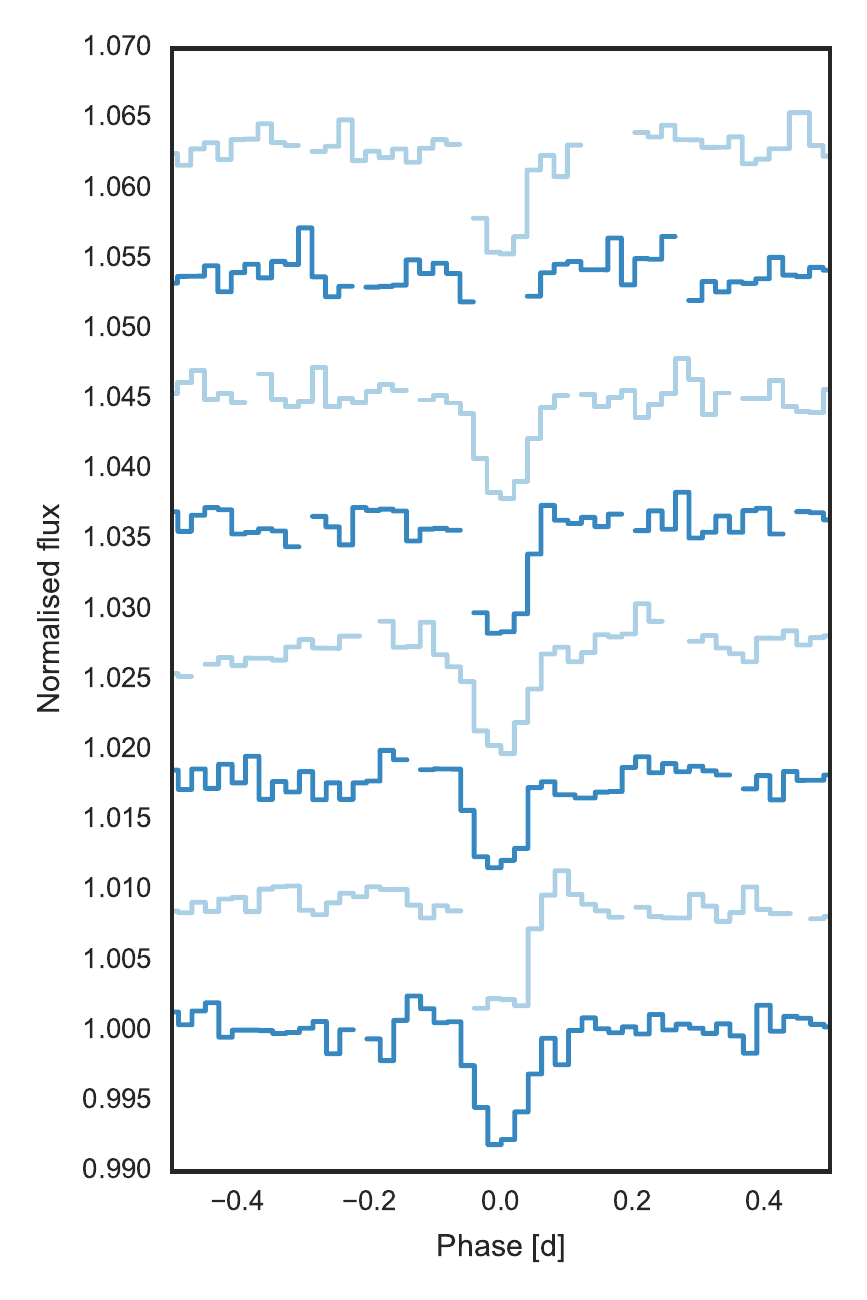}
    \caption{Individual transits in the \textsc{K2SC}-detrended PDC light curve of \pname. The transits are shown in chronological order from bottom to top, with a vertical offset applied for clarity.}
    \label{fig:K2_lc_indiv}
\end{figure}

While the PDC light curve is usually the version of choice to look for transit signals, the PDC pipeline sometimes removes true astrophysical signals on timescales $>20\,d$. Therefore, it is usually considered advisable to use the SAP light curve as a starting point for any detailed analysis of the transits and out-of-transit variability. However, after a visual comparison of the SAP and PDC light curves (both before and after modelling with \textsc{K2SC}), we opted to work with the PDC light curve throughout this paper. This decision was made because the SAP light curve is affected by long-term trends which are probably systematic, and because the \textsc{K2SC} modelling of the (shorter timescale) pointing-related systematics is less successful in the SAP than in the PDC light curve for this particular object. Figure~\ref{fig:K2_lc} shows the \ktwo\ PDC light curve for \pname\ along with the variability and systematics models from \textsc{K2SC}, as well as the light curve after correcting systematics only, and after also removing the variability model.  Figure~\ref{fig:K2_lc_indiv} shows the individual transits in more detail, using the detrended version of the light curve shown as version (e) in Figure~\ref{fig:K2_lc}. We estimated the 6-hour Combined Differential Photometric Precision (CDPP) of the fully detrended light curve as 274\,ppm (the CDPP is the standard measure of photometric precision on transit timescales for \emph{Kepler} light curves, see \citealt{2012PASP..124.1279C}).

\subsubsection{Transit detection}

The transits of \pname\ were identified independently by two methods. The first was visual inspection of the SAP and PDC light curves for all likely members of Praesepe observed by \ktwo\ (the transits are clearly visible in Fig.~\ref{fig:K2_lc}). The second was a systematic search for transits, which we performed for all the \textsc{K2SC}-detrended PDC light curves for Campaign 5 (after subtraction of both systematics and variability models), in which \pname\ was identified. The details of this search are given in \citet{Pope:2016}, who also report the full list of transit candidates. The initial search found transits with a period of $\sim 10.1$\,d, a depth of $\sim 7$ parts per thousand, and a duration of $\sim 3$\,h, consistent with a transiting companion somewhat smaller than Jupiter. A full analysis of the transits is presented in \S\ref{sec:transit_model}.

\subsubsection{Out-of-transit variability and rotation}

The light curve displays smooth, quasi-periodic variations with a peak-to-peak amplitude $\sim 2\%$, which we attribute to rotational modulation due to star spots. The \textsc{K2SC} pipeline detected these variations and automatically switched to using a quasi-periodic GP covariance function to model them. This covariance function is parametrized by a period, an amplitude, a periodic length scale (which controls the number of inflexions per period) and an evolutionary timescale. The best-fit values of the period and evolutionary timescale parameters were $25.5$\,d and $\sim 200$\,d, respectively. The former can be interpreted as an estimate of the host star's rotation period, while the latter suggests that the characteristic lifetime of the active regions present on the star's surface at the time of the observations was roughly 8 times the rotation period. Recent studies of rotation in Praesepe \citep{2011MNRAS.413.2218D,2011ApJ...740..110A,2014ApJ...795..161D} show a tight relationship between period and color from $J-K\sim0.25$ to $J-K \sim 0.8$, with periods tightly clustered around values ranging from $\sim 6$\,d at the bluer end to $\sim 10$\,d at the redder end. Redwards of $J-K \sim 0.8$, however, the rotation period  distribution becomes extremely broad, with periods ranging from 0.2 to $>30$\,d irrespective of color. With $J-K=0.84$, \pname\ lies in this second regime, and a rotation period of $\sim 25$\,d is thus consistent with, albeit at the longer end of, the expected range.

\subsection{Follow-up Light Curve Observations}
\label{sec:usno_lc}

In order to confirm the properties of the transit, we observed \pname\ on the night of UT 2016 February 26, using a thinned, back-side illuminated Tektronix 2048 CCD (``new2K") on the USNO, Flagstaff Station 40-inch telescope. A standard Kron-Cousins $I$-band filter was used with an exposure time of 120 sec (based on test exposures the previous night) and readout overhead of 51 sec, resulting in a net observing cadence of 171 sec. A total of 160 frames of \pname\ were obtained between UT 02:17 and UT 09:51. The frames were processed in real time with bias and flat-field frames obtained at the beginning of the night. Sky conditions were clear, although the seeing gradually worsened from about 1.2 arcsec FWHM at the beginning of the night to about 2.2 arcsec at the end of the exposure series. 

As a result of these observations, we obtain a clear detection of the transit, albeit at a lower photometric precision than from \ktwo.  The detection took place about 221 days after the last of the \ktwo\ observations, an offset of 22 orbital periods.  The USNO light curve is shown in conjunction with the \ktwo\ light curve in Figure~\ref{fig:LC_ecc_tlg}. In \S \ref{sec:transit_model} we show that incorporating the USNO light curve into the analysis gives a consistent solution with modelling the \ktwo\ light curve alone.

\subsection{Spectroscopic Observations and Radial Velocities}

We observed \pname\ with the Keck/HIRES spectrograph \citep{Vogt:1994} at six epochs: one in 1997 as part of another program, two in 2015, and three in 2016 as follow-up to the \ktwo\ light curve (see Table~\ref{tab:hires}).  All observations achieved spectral resolution $R>48,000$. The 1997 observation used a wavelength range of 6300\AA\ to 8725\AA, while the 2015 and 2016 observations used a range of 4800\AA\ to 9200\AA, except for the final one.  The images were processed and spectra extracted and calibrated using the \textsc{makee} software\footnote{http://www.astro.caltech.edu/$\sim$tb/makee/} written by Tom Barlow. The spectral type estimate for the star from these high-dispersion data is M3-M4. We infer the presence of a small amount of stellar chromospheric activity based on the fact that the H$\alpha$ line is completely filled in, with no apparent absorption or emission and thus no equivalent width measurement is possible.  The H$\beta$ line is weakly in emission with $EW(H\beta)=-0.45$ \AA.

From these spectra we obtained single-lined radial velocities of the host star.  Heliocentric velocities were measured in the four Keck/HIRES spectra by cross-correlating with radial velocity standards obtained at the same time as the program spectra.  The mean and standard deviations of the multi-order radial velocities are summarized in Table~\ref{tab:hires}.  The systemic radial velocity inferred from our four measurements is 35.34$\pm$0.17 km s$^{-1}$.  

The mean RV of the Praesepe cluster based on high resolution spectra of FGK members has been determined as 37.7 km s$^{-1}$ \citep{Barrado:1998}, and as $34.6 \pm 0.7$ km s$^{-1}$ \citep{Mermilliod:1999}.  The mean RV of \pname\ from the observations reported here is in between these values, thus we conclude that \pname\ is dynamically consistent with Praesepe cluster membership.

In addition, \citet{Barrado:1998} obtained medium-resolution spectra of a number of M dwarfs in Praesepe, including \pname\ (using the designation HSHJ 163).   For \pname, \citet{Barrado:1998} measure an RV of 32.4 km s$^{-1}$, which was consistent with Praesepe membership given the low RV accuracy from that paper, which was reported to be $\sim$7 km s$^{-1}$.  In Sec.~\ref{sec:properties} we use the six RV measurements reported here to set an upper limit on the mass of the transiting planet.

\begin{deluxetable}{ccccc}
\tablewidth{0pt}
\tabletypesize{\footnotesize}
\tablecolumns{5}
\tablecaption{RV observation of \pname\ with HIRES}
\tablehead{
    \colhead{Epoch} & \colhead{Spectral Range} & \colhead{SNR} &\colhead{RV Value} & \colhead{RV Error} \\
     UT Date & & & km s$^{-1}$ & km s$^{-1}$ \\
    }
\startdata
1997 12 06 15:02:32  & 6300 - 8725 \AA &  8 & 34.72 & 0.77 \\
2015 12 24 15:55:01  & 4800 - 9200 \AA & 21 & 34.24 & 0.33 \\
2015 12 29 16:12:14  & 4800 - 9200 \AA &  8 & 35.76 & 0.36 \\
2016 05 20 07:22:30  & 4800 - 9200 \AA & 28 & 36.18 & 0.35 \\
2016 12 22 14:55:18  & 4800 - 9200 \AA & 25 & 34.14 & 0.27 \\
2016 12 26 13:01:33  & 4100 - 8000 \AA &  9 & 36.72 & 1.11 \\
\enddata
\label{tab:hires}
\end{deluxetable}

\section{Results\label{sec:results}}

\subsection{Transit Model}
\label{sec:transit_model}

Since we do not have dynamical confirmation of this planet from RV detection of the signal, we have taken two separate approaches to the analysis of the photometric observations, presented in the next two subsections.  The consistent results of the two approaches provide additional evidence for the reliability of the overall system properties.

\subsubsection{Primary Analysis: Gaussian Process Transit Model}
\label{sec:GP-transit}

The \ktwo\ light curve of \pname\ is shown in Figure \ref{fig:K2_lc} and displays evolving starspot modulation with an amplitude of $\sim$2\% and period $\sim$24 days. To account for the effect of this stellar variability across each transit, we simultaneously model the out-of-transit variations at the same time as fitting for the transit parameters. We therefore model the full light curve as the sum of a Gaussian process plus transit model (hereafter \emph{GP-transit} model). The transit component is based on the model presented in \citet{Irwin:2011}, which is a modified version of the (JKT)EBOP family of models, but which uses the analytic method of \citet{Mandel:2002} for the quadratic limb darkening law. The Gaussian process (GP) component, which is used to describe the out-of-transit variability, is based on the {\tt george} package \citep{Ambikasaran:2014,georgeDFM2014}. For the \ktwo\ light curve, the GP component is a quasi-periodic exponential-sine-squared kernel, which is essentially a periodic kernel that is allowed to evolve over time, i.e. mimicking evolving starspot modulation.

In addition to the \ktwo\ discovery light curve we use the USNO light curve described in \S \ref{sec:usno_lc}. The single-night observation is not long enough to clearly see the evolving starspot modulation displayed in the \ktwo\ light curve. Instead, the dominant variations display a relatively rough behavior over short ($\lesssim$1 hour) timescales. Accordingly, we opt for a Matern-3/2 kernel in our GP model, as its covariance properties are more suited to the observed behavior.

We simultaneously modeled the \ktwo\ and USNO light curves using our \emph{GP-transit} model, with the parameter space explored through the Affine Invariant Markov chain Monte Carlo (MCMC) method, as implemented in {\tt emcee} \citep{Foreman-Mackey:2013}. This model will be presented in detail in Gillen et al. (\emph{in prep}). We ran the MCMC for 50\,000 steps with each of 144 `walkers' and derived parameter distributions from the last 25\,000 steps of each chain (after thinning each chain from inspection of the autocorrelation lengths for each parameter). In addition, the \ktwo\ model was supersampled to 1 min cadence. 

Since we do not have strong prior information about the eccentricity of this orbit, we opted to apply four different models to this system. These models are presented in Table \ref{table:priors} and differ only in their eccentricity, effective temperature and surface gravity priors, the latter two giving different priors on the stellar density: 
A) an eccentric orbit with a stellar density prior from the empirical relations of \citet{Mann:2016} (see \S \ref{sec:SED}), 
B) an eccentric orbit with a stellar density prior from the SED fitting (again see \S \ref{sec:SED}), 
C) an eccentric orbit with an uninformative stellar density prior, 
and D) a circular orbit with an uninformative stellar density prior. 
Model A is our main model. We then test this stellar density prior first by relaxing it slightly in model B and then relaxing it fully with an uninformative prior in model C. Finally, in model D, we keep the uninformative stellar density prior but force the orbit to circular to investigate the effect of eccentricity.
 
\begin{figure}[htb!]
   \centering
   \includegraphics[width=\linewidth]{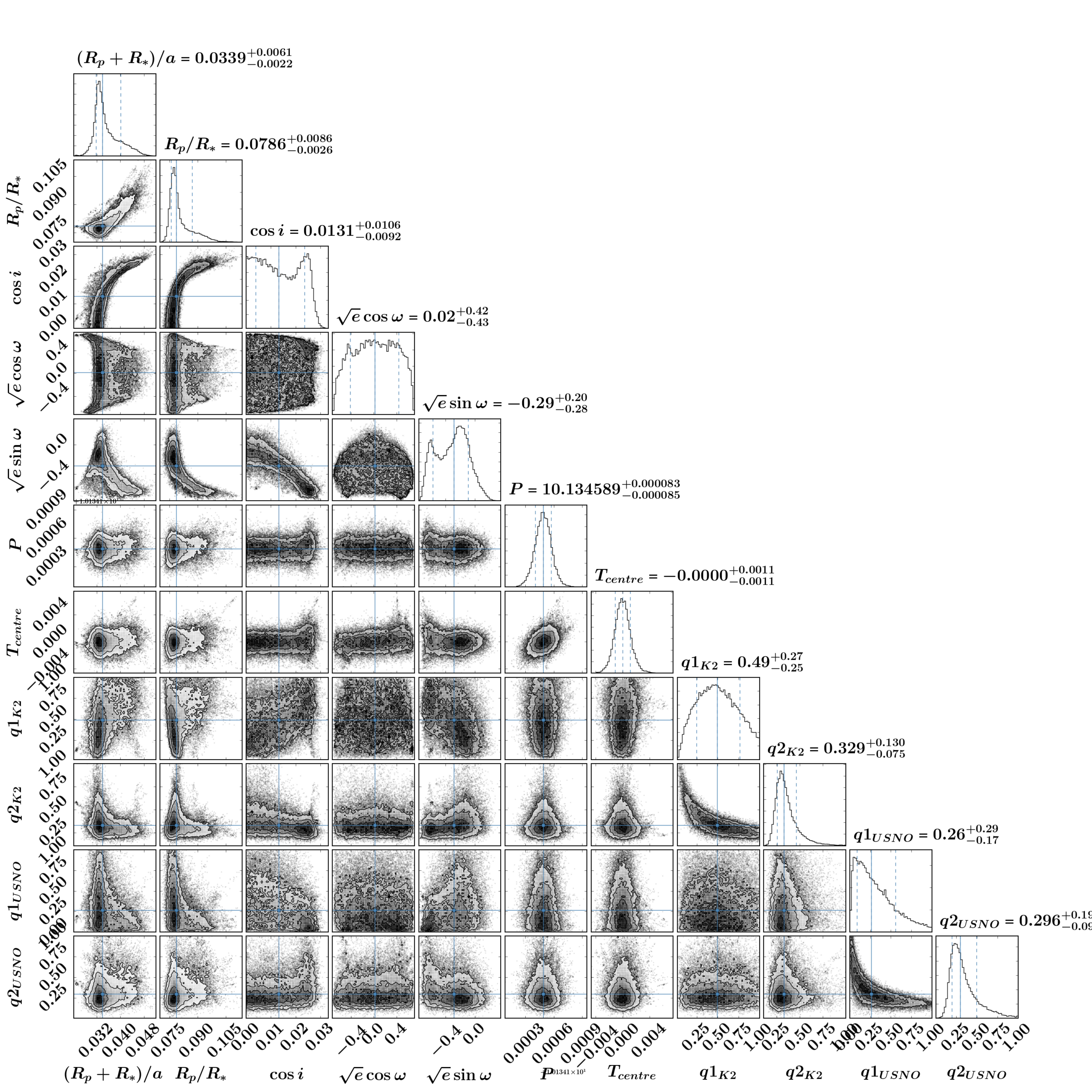}
   \caption{2D contours and 1D histograms of the MCMC chains for the \emph{GP-transit} model A, showing the transit parameters only of both the \ktwo\ and USNO light curve models. On the 2D plots, blue squares show the median values and black contours represent 1-, 2- and 3-$\sigma$ confidence intervals. On the histograms, solid and dotted vertical lines show the median and $\pm 1 \sigma$ intervals, respectively.  For limb-darkening we use the method of \citet{Kipping:2013}, and we quote the `q' parameters from that formulation here.  Figure created using the \textsc{corner.py} code from\citet{ForemanMackey2016}
   }
   \label{fig:LC_conf_ecc_tlg}
\end{figure}

The effective temperature and surface gravity priors in models A and B naturally give rise to priors on the limb darkening coefficients.
These were computed using the LDTK toolkit \citep{Parviainen2015pt}, which allows us to propagate the uncertainties in these stellar parameters (and the metallicity $Z=0.1\pm0.1$) through the PHOENIX stellar atmosphere models \citep{Husser:2013} into our coefficients. 
The uncertainties on the limb darkening coefficients were then inflated by a factor of 50 to account for differences between model grids. In the light curve modelling, the limb darkening parameterisation follows the triangular sampling method of \citet{Kipping:2013} (see that paper for the relationship between the  `q' parameters plotted in Fig \ref{fig:LC_conf_ecc_tlg} and standard quadratic limb darkening coefficients.) Uninformative priors were used on all other transit parameters.
 
Using models A--D we can explore the information content of the data, how our priors effect our results, and the confidence with which we can characterise this system. The individual models are presented and compared below.

\begin{table}[htb!]
 \centering  
 \scriptsize
\caption{Priors on System Parameters for Analysis in \S \ref{sec:GP-transit}
 \label{table:priors}}
\begin{tabular}{lcccc}  
 \noalign{\smallskip} \noalign{\smallskip} \hline  \hline \noalign{\smallskip}  
 Case  &   Eccentricity  &  Stellar density (g~cm$^{-3}$)   &   Effective temperature (K)   &  $\log g$   \\   
 \noalign{\smallskip} \hline \noalign{\smallskip} \noalign{\smallskip}  

A (eccentric, empirically-constrained SD*)      & $\mathcal{U}(0, 1.0)$        & $7.266\,^{+1.877}_{-1.431}$        &      $\mathcal{N}(\mu=3350, \sigma=50)$     &   $\mathcal{N}(\mu=4.79, \sigma=0.08)$      \\  [1.ex] 
B (eccentric, SED-constrained SD)               & $\mathcal{U}(0, 1.0)$        & $3.703\,^{+8.035}_{-2.534}$        &      $\mathcal{N}(\mu=3350, \sigma=50)$     &   $\mathcal{N}(\mu=4.5, \sigma=0.5)$      \\  [1.ex] 
C (eccentric, unconstrained SD)                   & $\mathcal{U}(0, 1.0)$        & $\mathcal{U}(0, 160)$       & ---       & ---  \\  [1.ex]
D (circular, unconstrained SD)                      & $\delta$                             & $\mathcal{U}(0, 160)$       & ---       & ---  \\  [1.ex]

 \noalign{\smallskip} \noalign{\smallskip}\noalign{\smallskip}  
 \hline  
 \end{tabular}  
\begin{flushleft}
\footnotesize * SD = stellar density \\
\footnotesize $\mathcal{U}(a, b)$ - uniform density with a minimum $a$ and maximum $b$ \\
\footnotesize $\mathcal{N}(\mu, \sigma)$ - normal density with mean $\mu$ and standard deviation $\sigma$ \\
\footnotesize $\delta$ - Dirac's delta function.
\end{flushleft}
\end{table}

\subsubsubsection{Individual models: investigating the effect of eccentricity and assumed stellar density}

Given that the planet orbital period lies close to an integer multiple of the \ktwo\ cadence, and that the USNO light curve is of modest precision, the eccentricity of the orbit is unlikely to be well-constrained from these data alone. As we believe the empirically-determined stellar density constraint to be well-founded, we opt to use model A (eccentric orbit with a stellar density constraint from empirical relations) as the main model discussed here, but also present results from the other three for comparison and to investigate the effect of our assumptions.

\begin{figure}[tb!]
   \centering
   \includegraphics[width=0.9\linewidth]{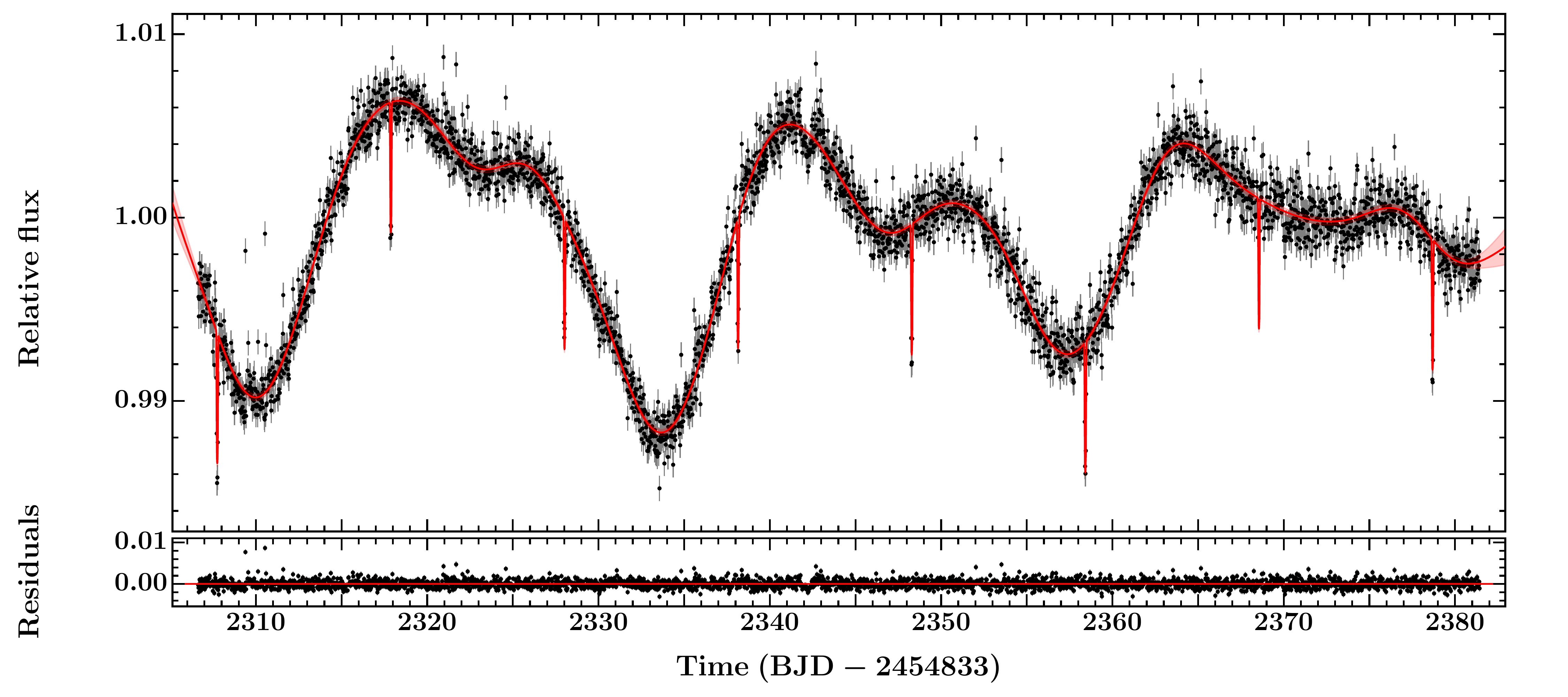}
   \includegraphics[width=0.65\linewidth]{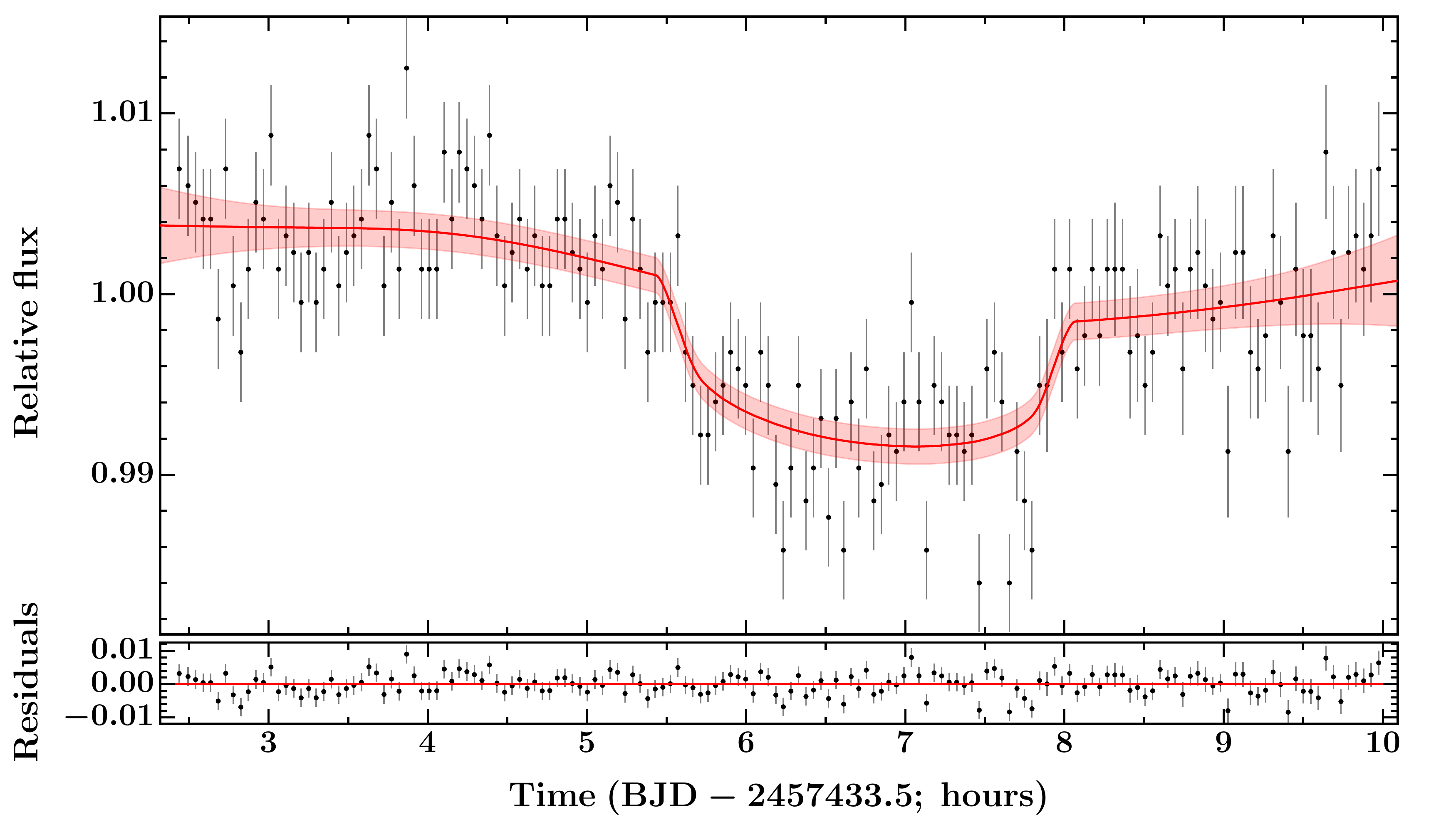} 
   \caption{\ktwo\ (top) and USNO (bottom) light curves of \pname\ (black) with the \emph{GP-transit} model A (red). The \ktwo\ light curve here is the same as the systematics-corrected PDC version shown in Fig \ref{fig:K2_lc}. The red line and pink shaded region show the mean and 2-sigma confidence interval of the predictive posterior distribution of the \emph{GP-transit} model. The USNO observations cover the transit 22 orbital periods later than the last event seen in the \ktwo\ data.}
   \label{fig:LC_ecc_tlg}
\end{figure}

\begin{figure}[htb!]
   \centering
   \includegraphics[width=0.49\linewidth]{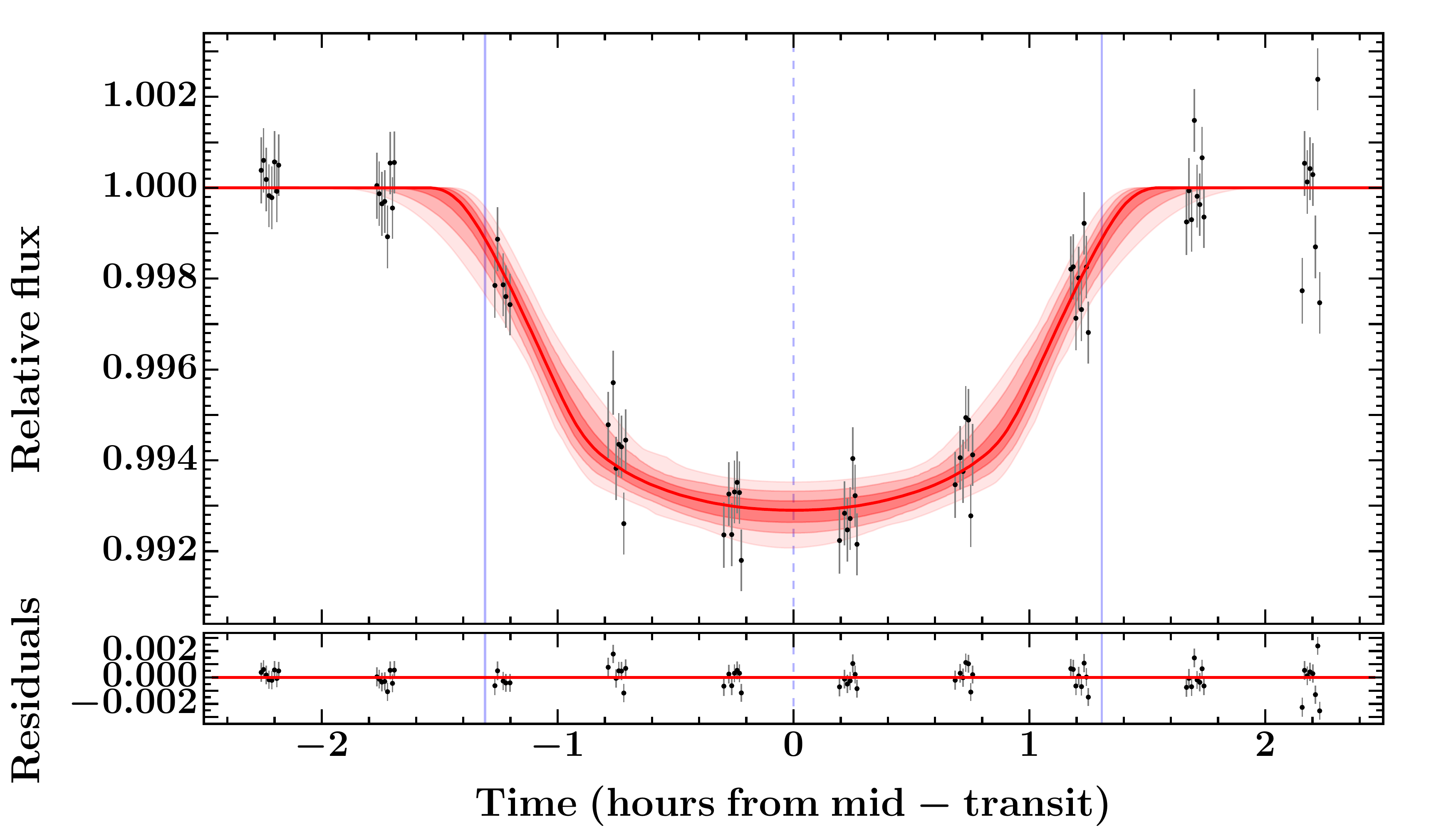}
   \includegraphics[width=0.49\linewidth]{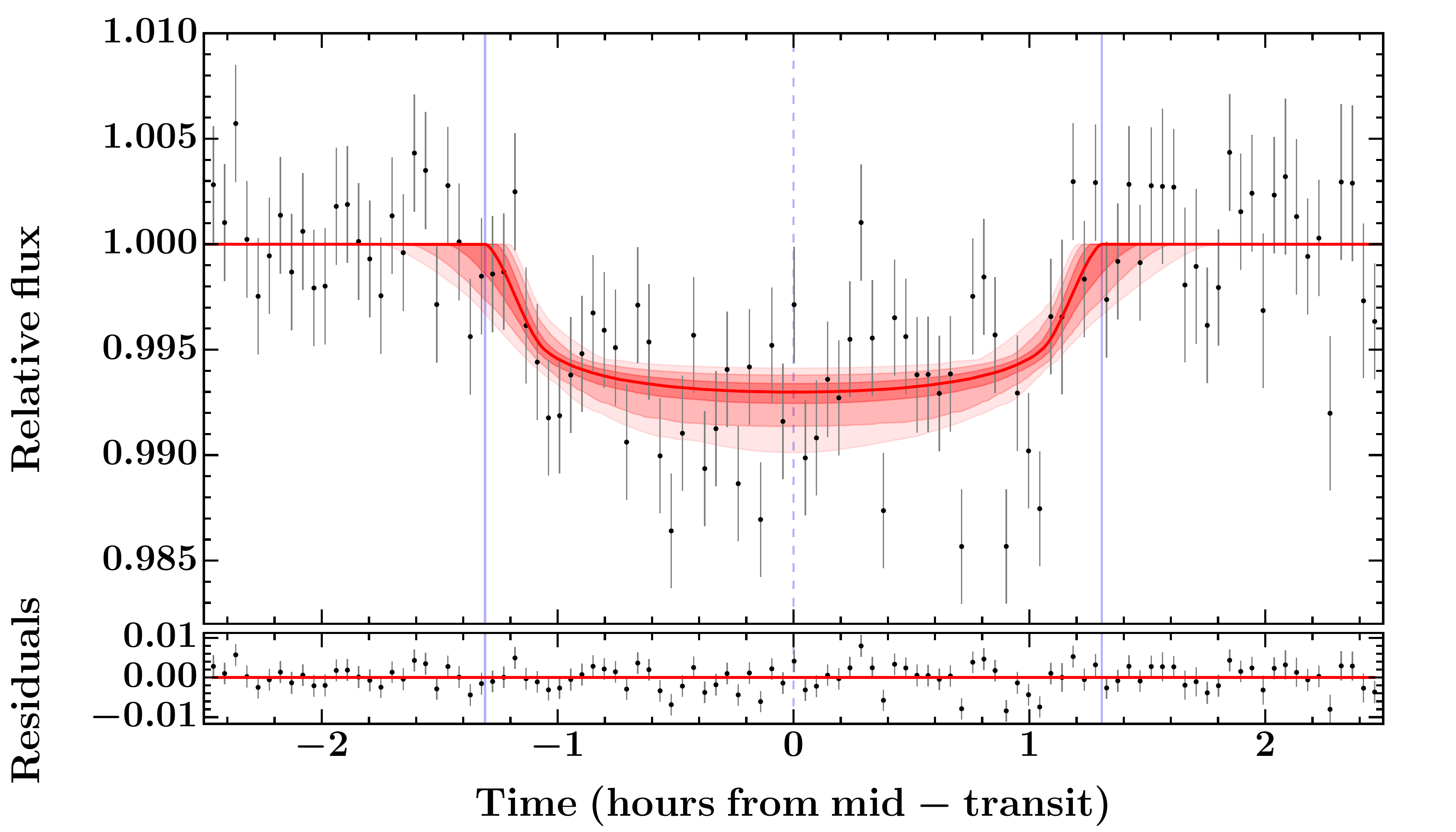}
   \includegraphics[width=0.49\linewidth]{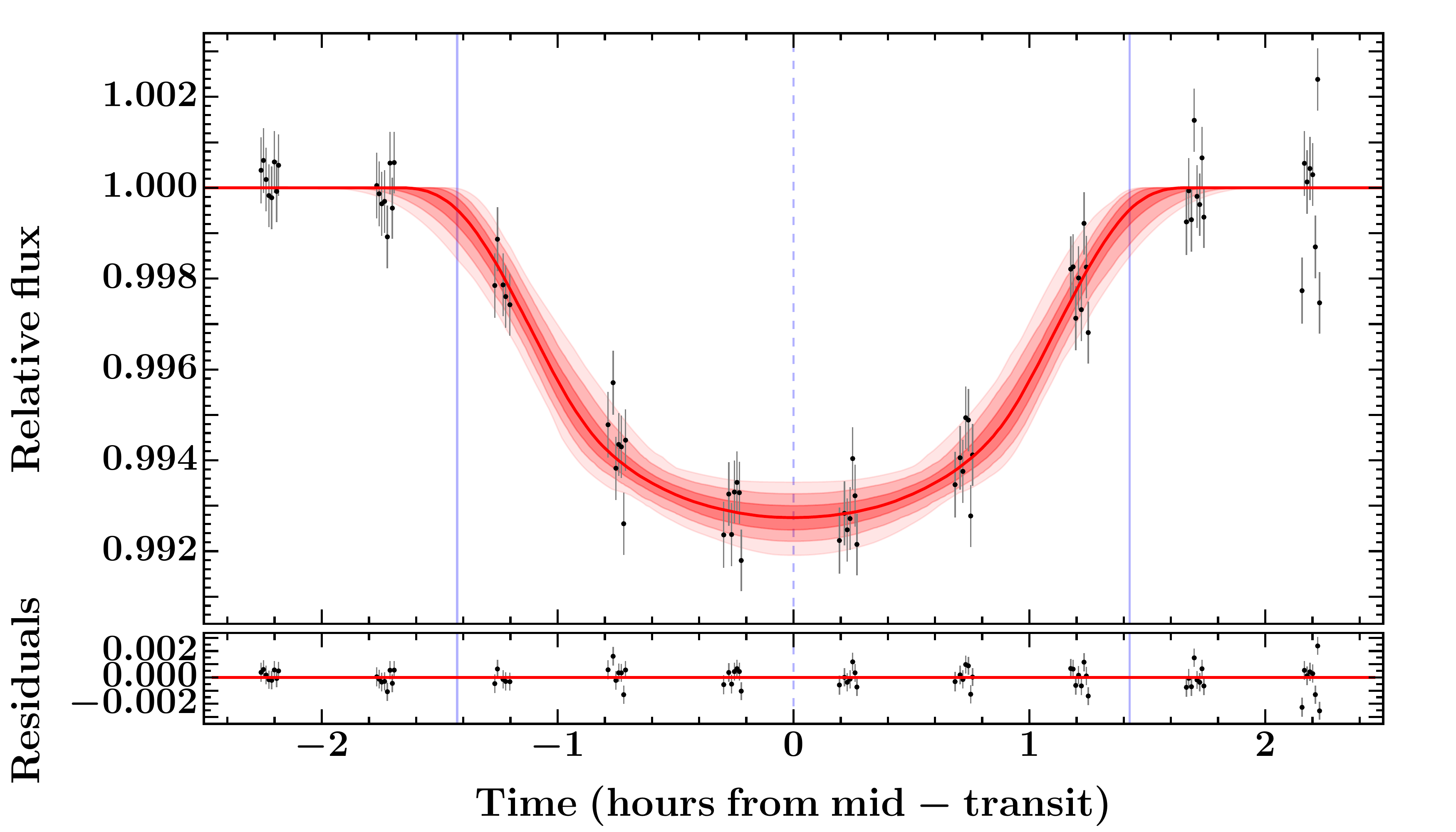}
   \includegraphics[width=0.49\linewidth]{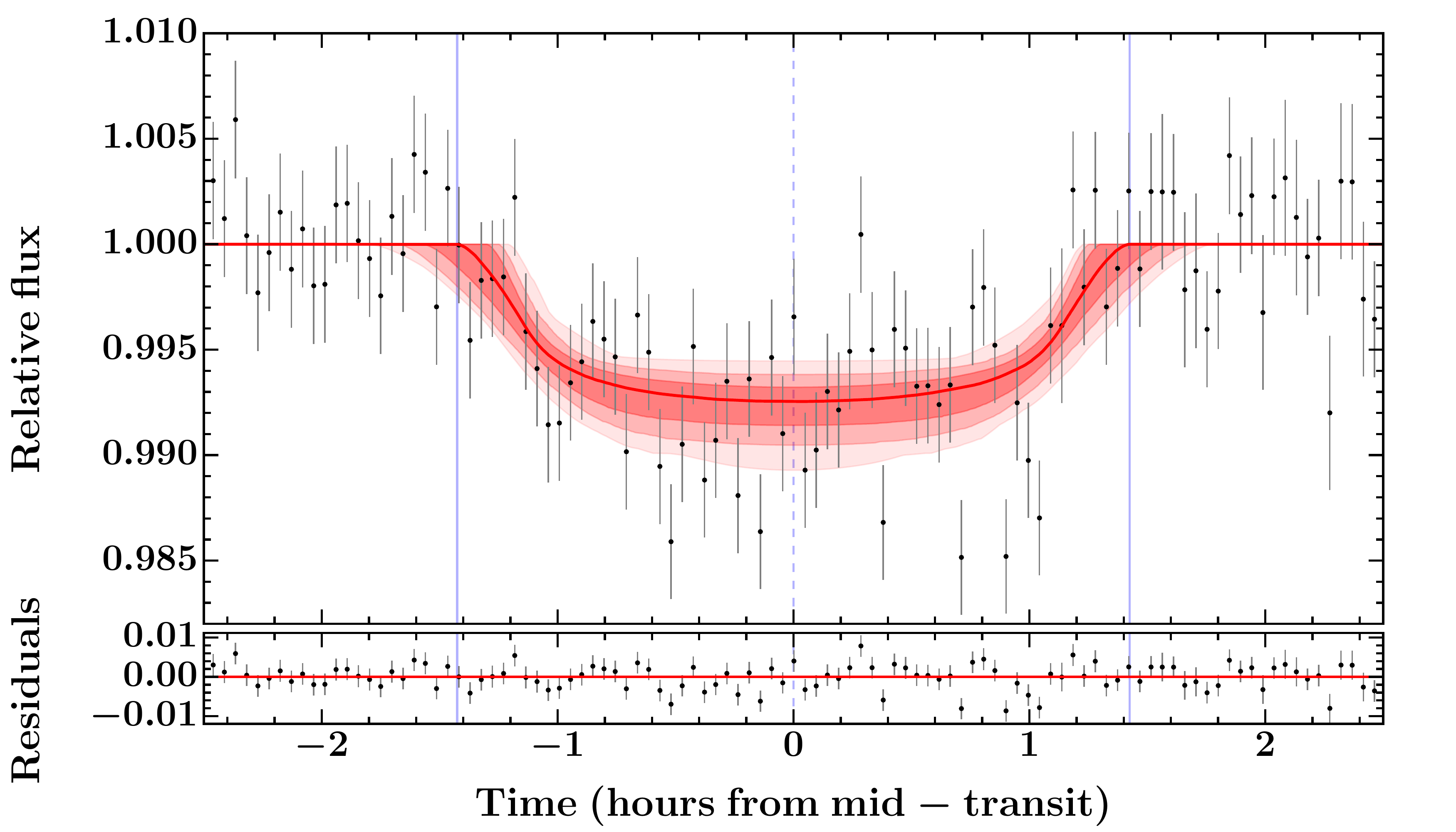}
   \includegraphics[width=0.49\linewidth]{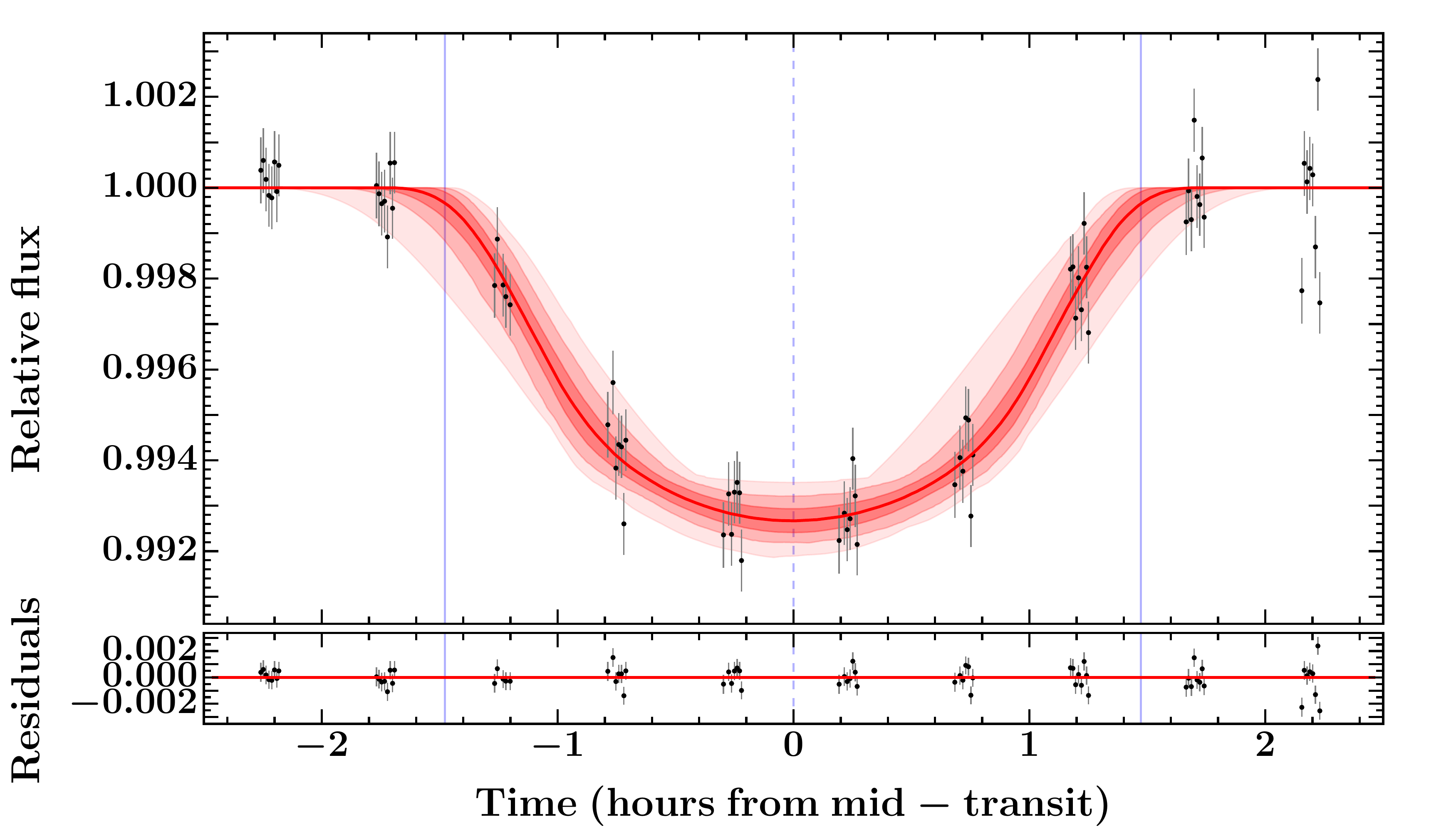}
   \includegraphics[width=0.49\linewidth]{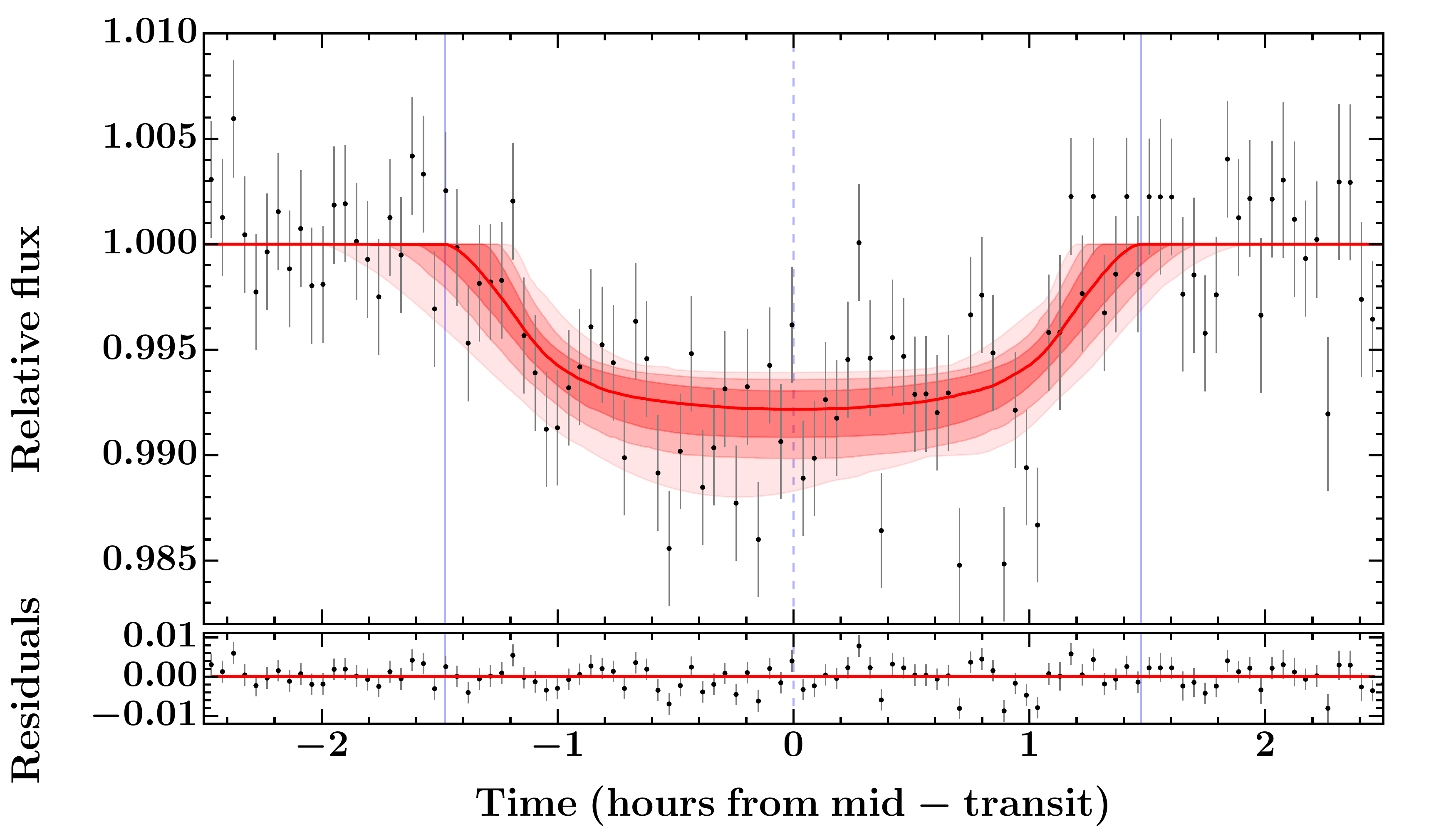}
   \includegraphics[width=0.49\linewidth]{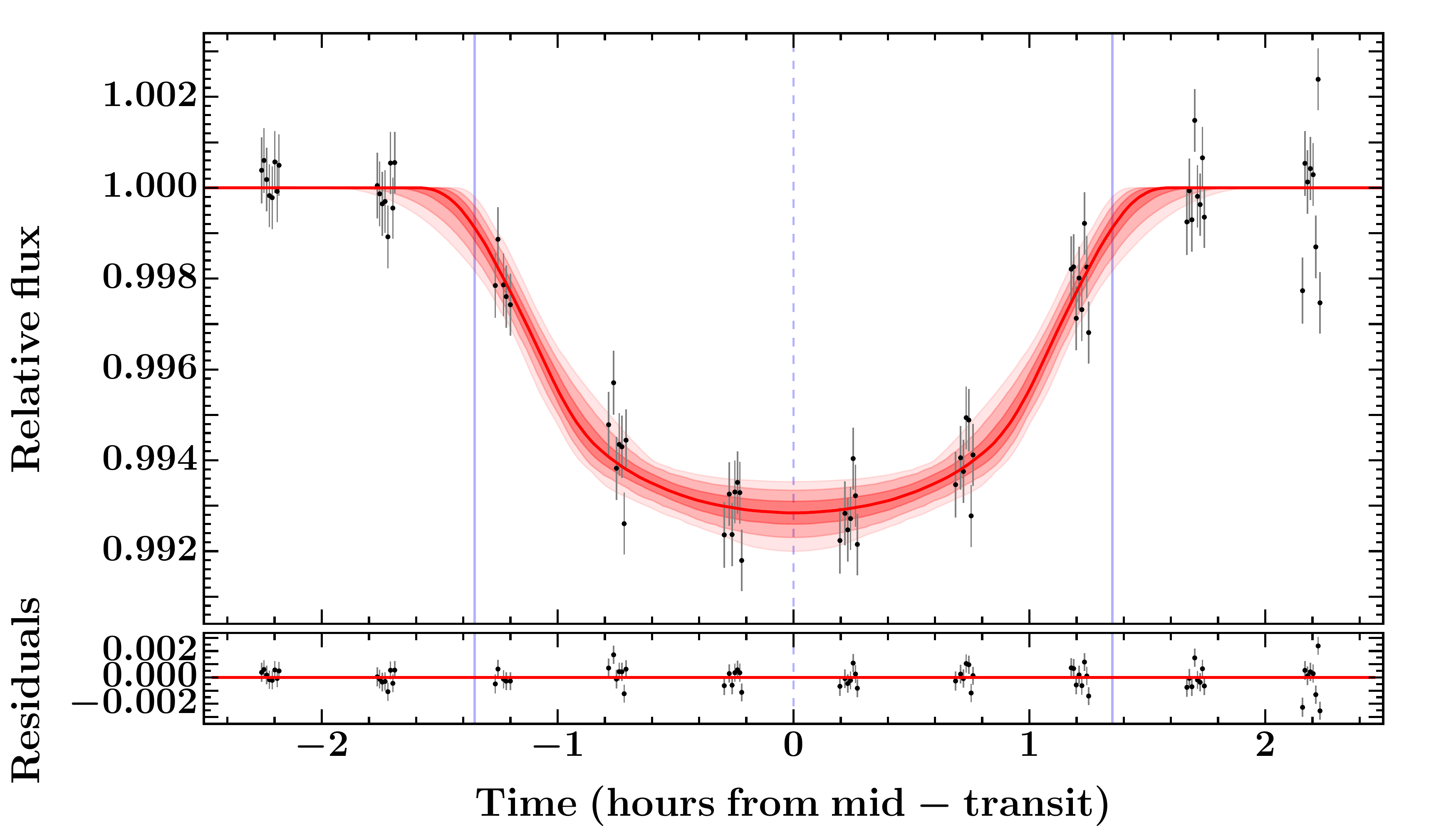}    
   \includegraphics[width=0.49\linewidth]{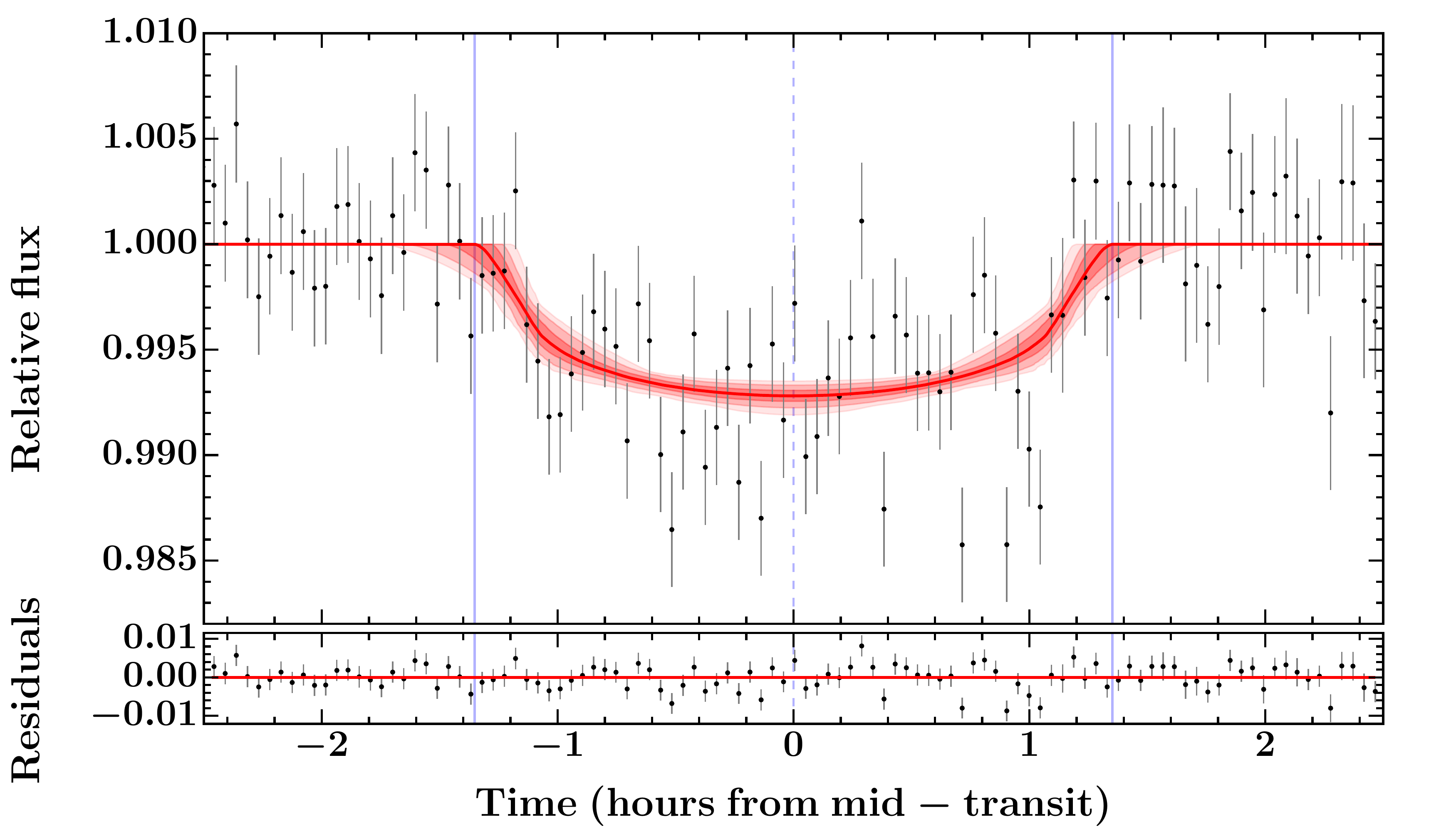}  
   \caption{Phase-folded \ktwo\ (\emph{left column}) and USNO (\emph{right column}) light curves, zoomed in around transit (black points) showing the \emph{GP-transit} model fits (red). The light curves have been detrended with respect to the GP models (i.e. stellar variability and large-scale systematics). The red line and pink shaded regions indicate the median and 1-, 2- and 3-sigma confidence regions of the \emph{GP-transit} model. The vertical dashed and solid blue lines represent the transit centre and first and fourth contacts, respectively. The \ktwo\ model has been integrated to the 30-min observational cadence, whereas the USNO model is not integrated given the short exposure times of the USNO data. \emph{Rows, top-to-bottom, models A--D}: model A (eccentric orbit, stellar density prior from empirical relations), model B (eccentric orbit, stellar density prior from the SED fitting), model C (eccentric orbit, uninformative stellar density prior) and model D (circular orbit, uninformative stellar density prior).}
   \label{fig:LC_K2_USNO_phase}
\end{figure}

Figure \ref{fig:LC_ecc_tlg} shows the full \ktwo\ and USNO light curves along with the \emph{GP-transit} model A. The model is an acceptable fit the the large scale structure of both light curves. In the \ktwo\ case, the GP model predicts a stellar rotation period of 23.8 days and is able to reproduce the evolving shape of the modulation. For the USNO light curve, the GP favors a smooth model for the out-of-transit variations due to the observational uncertainties, even though apparent correlations can be seen in the residuals. The significance of these variations (as well as lower level ones present in the \ktwo\ light curve) are investigated at the end of this section.
The top row of Figure \ref{fig:LC_K2_USNO_phase} shows the model A transit fits to both the \ktwo\ and USNO light curves (left and right, respectively). In the \ktwo\ panel, the effect of the orbital period being close to an integer multiple of the cadence can be seen.
Table \ref{tab:lc_model_tab} presents the parameters of model A (leftmost results column) and Figure \ref{fig:LC_conf_ecc_tlg} presents the 2D contours and 1D histograms of the transit parameter MCMC chains from which the posteriors reported in Table \ref{tab:lc_model_tab} are derived. 
There are significant correlations between the parameters of interest, as expected given we only have photometric data of the transit (i.e. no secondary eclipse or radial velocities). Allowing an eccentric orbit drives most of the observed correlations: fixing the orbit to circular removes the  correlations of interest, bar those between the cosine of the inclination ($\cos i$) and the radius sum ($(R_{p}+R_{*})/a$). The correlations arise from the fact that small changes in a given parameter can be effectively masked by corresponding changes in one or more others without significant modification to the transit shape.

 \begin{figure}[htb!]
   \centering
   \includegraphics[width=0.7\linewidth]{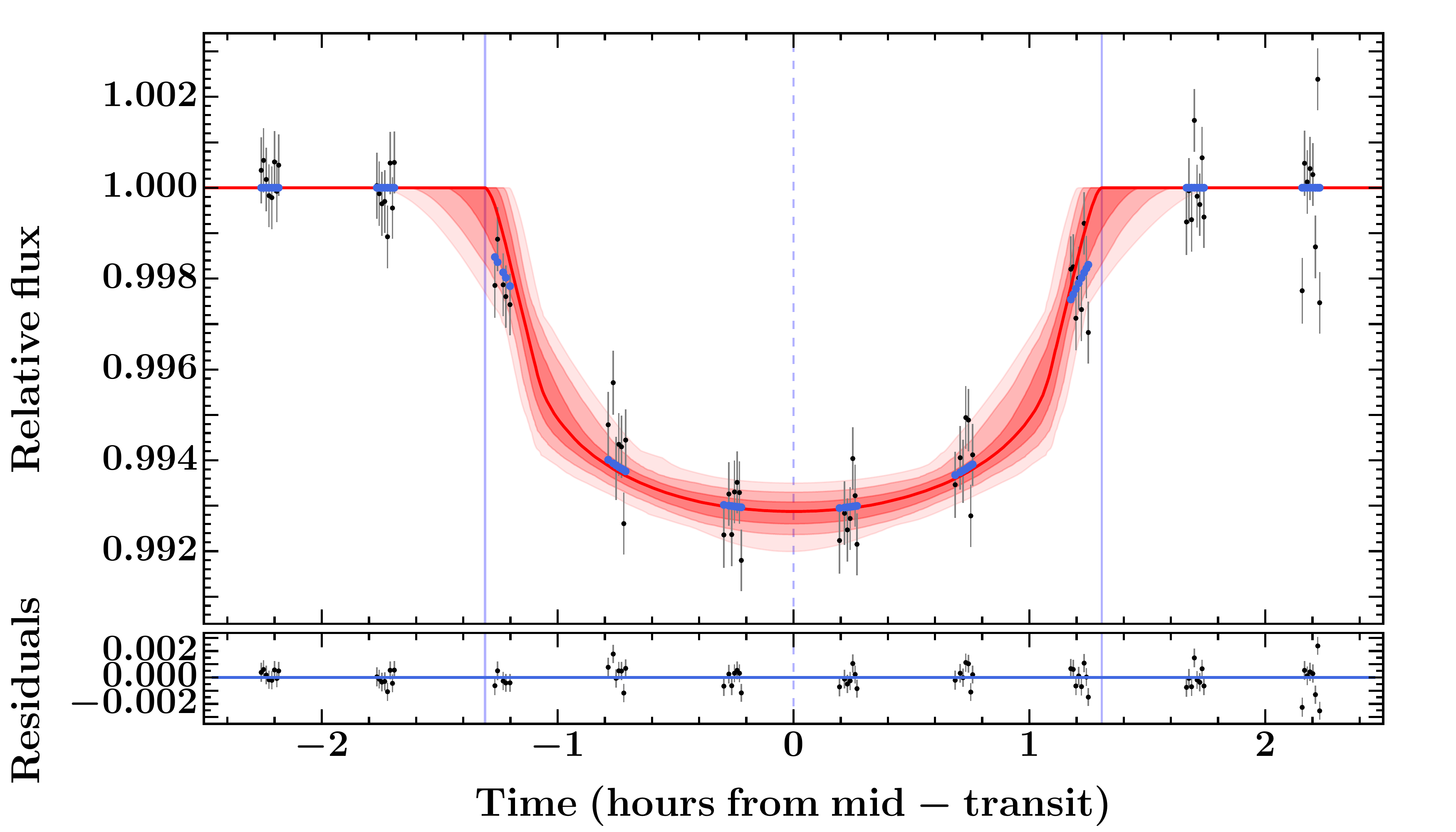}
   \caption{Unintegrated \ktwo\ transit model for Model A. The light curve has been detrended with respect to the GP model. The red line and pink shaded regions indicate the median and 1-, 2- and 3-sigma confidence regions of the unintegrated \emph{GP-transit} model. The blue points show the integrated model predictions for the times of observation.  The vertical dashed and solid blue lines represent the transit centre and first and fourth contacts, respectively.}
   \label{fig:LC_model_nointeg}
\end{figure}

\begin{landscape}  
 \begin{table*}[htbp!]  
 \centering  
 \small  
 \caption[Parameters of LC model for EPIC 211916756]{Parameters of the different light curve models applied to \pname.}
 \label{tab:lc_model_tab}  
 \begin{tabular}{l l l l l l l }  
 \noalign{\smallskip} \noalign{\smallskip} \hline  \hline \noalign{\smallskip}  
 Parameter  &   Symbol  &  Unit  & \multicolumn{4}{c}{Value} \\   
     &        &            &    \emph{\,\,\,\,\,Model A}    &    \emph{\,\,\,\,\,Model B}    &    \emph{\,\,\,\,\,Model C}    &    \emph{\,\,\,\,\,Model D} \\ 
 \noalign{\smallskip} \hline \noalign{\smallskip} \noalign{\smallskip}  
 \multicolumn{7}{c}{\emph{Transit parameters}} \\   
 \noalign{\smallskip} \noalign{\smallskip}   
 Sum of radii    &    $(R_{\rm{*}} + R_{\rm{p}})/ a$    &        &    $\,\,\,\,\,0.0339\,^{+0.0061}_{-0.0022}$    &    $\,\,\,\,\,0.050\,^{+0.020}_{-0.013}$    &    $\,\,\,\,\,0.069\,^{+0.026}_{-0.020}$    &    $\,\,\,\,\,0.0403\,^{+0.0175}_{-0.0062}$    \\  [1.1ex]  
 Radius ratio    &    $R_{\rm{p}} / R_{\rm{*}}$    &       &    $\,\,\,\,\,0.0786\,^{+0.0086}_{-0.0026}$    &    $\,\,\,\,\,0.0852\,^{+0.0097}_{-0.0084}$    &    $\,\,\,\,\,0.0887\,^{+0.0097}_{-0.0113}$    &    $\,\,\,\,\,0.0794\,^{+0.0084}_{-0.0040}$             \\  [1.1ex]  
 Orbital inclination & $i$ & $^{\circ}$        &    $\,\,\,\,\,89.25\,^{+0.53}_{-0.61}$    &    $\,\,\,\,\,88.0 \pm 1.6$    &    $\,\,\,\,\,86.4 \pm 4.2$    &    $\,\,\,\,\,88.82\,^{+0.82}_{-1.30}$    \\  [1.1ex]  
 Orbital period    &    $P$    &    days         &    $\,\,\,\,\,10.134589 \pm 0.000084$    &    $\,\,\,\,\,10.134590 \pm 0.000098$    &    $\,\,\,\,\,10.134603 \pm 0.000102$    &    $\,\,\,\,\,10.134588 \pm 0.000094$    \\  [1.1ex]  
 Time of transit centre    &    $T_{\rm{centre}}$    &    BJD         &    $\,\,\,\,\,2457282.6279 \pm 0.0011$    &    $\,\,\,\,\,2457282.6279 \pm 0.0015$    &    $\,\,\,\,\,2457282.6281 \pm 0.0020$    &    $\,\,\,\,\,2457282.6278 \pm 0.0012$    \\  [1.1ex]  
     &    $\sqrt{e} \cos \omega$    &         &    $\,\,\,\,\,0.02 \pm 0.43$    &    $\,\,\,\,\,0.01 \pm 0.52$    &    $-0.03 \pm 0.6$    &        \\  [1.1ex]  
     &    $\sqrt{e} \sin \omega$    &         &    $-0.29\,^{+0.20}_{-0.28}$    &    $-0.04 \pm 0.42$    &    $\,\,\,\,\,0.28\,^{+0.38}_{-0.50}$    &        \\  [1.1ex]  
 Eccentricity    &    $e$    &         &    $\,\,\,\,\,0.24\,^{+0.27}_{-0.18}$    &    $\,\,\,\,\,0.27\,^{+0.27}_{-0.19}$    &    $\,\,\,\,\,0.46\,^{+0.21}_{-0.30}$    &        \\  [1.1ex]  
 Longitude of periastron    &    $\omega$    &    $^{\circ}$         &    $-76.8\,^{+53.0}_{-69.7}$    &    $-10.0 \pm 130.0$    &    $\,\,\,\,\,58.0 \pm 75.0$    &        \\  [1.1ex]  
 \noalign{\smallskip} \noalign{\smallskip}K2 Linear LD coefficient    &    $u_{\rm{K2}}$    &         &    $\,\,\,\,\,0.456 \pm 0.082$    &    $\,\,\,\,\,0.67 \pm 0.26$    &    $\,\,\,\,\,0.88 \pm 0.38$    &    $\,\,\,\,\,0.60 \pm 0.27$    \\  [1.1ex]  
 K2 Non-linear LD coefficient    &    $u'_{\rm{K2}}$    &         &    $\,\,\,\,\,0.24 \pm 0.20$    &    $\,\,\,\,\,0.09 \pm 0.29$    &    $-0.11 \pm 0.39$    &    $\,\,\,\,\,0.07 \pm 0.34$    \\  [1.1ex]  
 USNO Linear LD coefficient    &    $u_{\rm{USNO}}$    &         &    $\,\,\,\,\,0.304 \pm 0.079$    &    $\,\,\,\,\,0.27 \pm 0.22$    &    $\,\,\,\,\,0.28\,^{+0.31}_{-0.20}$    &    $\,\,\,\,\,0.60 \pm 0.27$    \\  [1.1ex]  
 USNO Non-linear LD coefficient    &    $u'_{\rm{USNO}}$    &         &    $\,\,\,\,\,0.21 \pm 0.22$    &    $\,\,\,\,\,0.16\,^{+0.31}_{-0.22}$    &    $\,\,\,\,\,0.13\,^{+0.32}_{-0.23}$    &    $\,\,\,\,\,0.07 \pm 0.34$    \\  [1.1ex]  
 \noalign{\smallskip} \noalign{\smallskip} \noalign{\smallskip} \noalign{\smallskip}  \noalign{\smallskip}  
 \multicolumn{7}{c}{\emph{Out-of-transit variability parameters}} \\  
 \noalign{\smallskip} \noalign{\smallskip}  
 K2 Amplitude    &    $A_{\rm{K2}}$    &    \%         &    $\,\,\,\,\,0.230\,^{+0.058}_{-0.042}$    &    $\,\,\,\,\,0.226\,^{+0.056}_{-0.042}$    &    $\,\,\,\,\,0.222\,^{+0.057}_{-0.041}$    &    $\,\,\,\,\,0.226\,^{+0.054}_{-0.041}$    \\  [1.1ex]  
 K2 Timescale of SqExp term    &    $l_{\rm{SE ~ K2}}$    &    \%         &    $\,\,\,\,\,156.2 \pm 3.4$    &    $\,\,\,\,\,156.2 \pm 3.3$    &    $\,\,\,\,\,155.6 \pm 3.3$    &    $\,\,\,\,\,156.3 \pm 3.3$    \\  [1.1ex]  
 K2 Scale factor of ExpSine2 term    &    $\Gamma_{\rm{ESS ~ K2}}$    &    days         &    $\,\,\,\,\,0.484 \pm 0.090$    &    $\,\,\,\,\,0.491\,^{+0.092}_{-0.086}$    &    $\,\,\,\,\,0.494 \pm 0.092$    &    $\,\,\,\,\,0.490 \pm 0.088$    \\  [1.1ex]  
 K2 Period of ExpSine2 term    &    $P_{\rm{ESS ~ K2}}$    &    days         &    $\,\,\,\,\,23.83 \pm 0.29$    &    $\,\,\,\,\,23.84 \pm 0.28$    &    $\,\,\,\,\,23.85 \pm 0.31$    &    $\,\,\,\,\,23.83 \pm 0.27$    \\  [1.1ex]  
 K2 White noise term    &    $\sigma_{\rm{K2}}$    &    \%         &    $\,\,\,\,\,1.371 \pm 0.017$    &    $\,\,\,\,\,1.371 \pm 0.018$    &    $\,\,\,\,\,1.371 \pm 0.017$    &    $\,\,\,\,\,1.371 \pm 0.017$    \\  [1.1ex]  
 \noalign{\smallskip} \noalign{\smallskip}USNO Amplitude    &    $A_{\rm{USNO}}$    &    \%         &    $\,\,\,\,\,0.36 \pm 0.18$    &    $\,\,\,\,\,0.32 \pm 0.18$    &    $\,\,\,\,\,0.31 \pm 0.20$    &    $\,\,\,\,\,0.37 \pm 0.17$    \\  [1.1ex]  
 USNO Timescale    &    $l_{\rm{USNO}}$    &    days         &    $\,\,\,\,\,5.0\,^{+2.6}_{-2.0}$    &    $\,\,\,\,\,5.9 \pm 2.7$    &    $\,\,\,\,\,6.4 \pm 2.6$    &    $\,\,\,\,\,5.0 \pm 2.4$    \\  [1.1ex]  
 USNO White noise term    &    $\sigma_{\rm{USNO}}$    &    \%         &    $\,\,\,\,\,1.167 \pm 0.071$    &    $\,\,\,\,\,1.175\,^{+0.079}_{-0.069}$    &    $\,\,\,\,\,1.175\,^{+0.075}_{-0.068}$    &    $\,\,\,\,\,1.173\,^{+0.071}_{-0.065}$    \\  [1.1ex]  
 \noalign{\smallskip} \noalign{\smallskip}\noalign{\smallskip}  
 \hline  
 \end{tabular}  
 \begin{list}{}{}  
 \item[* LD = limb darkening]  
 \end{list}  
 \end{table*}  
 \end{landscape}

We now compare the four different models to investigate the effect of our assumptions about the eccentricity and stellar density. The four rows of Figure \ref{fig:LC_K2_USNO_phase} show (in descending order) the fits of models A, B, C and D. All models are able to reproduce the observed transit shapes within the observational uncertainties, even though they have different shapes and durations.  
These transit models are integrated to the \ktwo\ cadence. To highlight the effect of this observational cadence on the depicted transit models, Figure \ref{fig:LC_model_nointeg} shows the unintegrated transit of Model A (red line and shaded regions) and the integrated model predictions for the times of observation (blue points).

\begin{figure}[htb!]
   \centering
   \includegraphics[width=\linewidth]{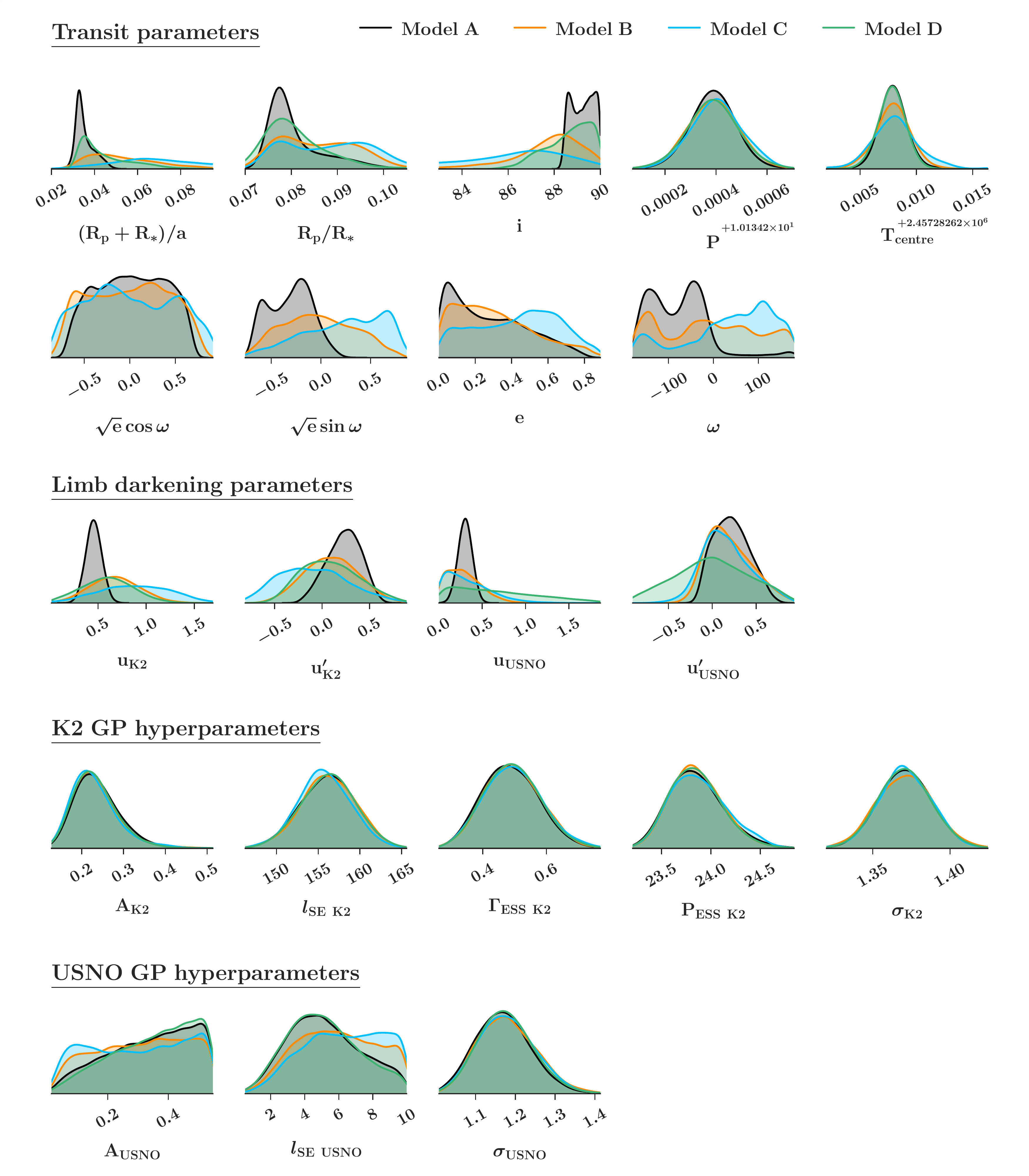}
   \caption{Posterior parameter distributions of the four \emph{GP-transit} models: A (black; main model), B (orange), C (blue) and D (green). The transit parameters, limb darkening coefficients and GP hyperparameters are grouped from top to bottom.}
   \label{fig:LC_model_comp}
\end{figure}

Figure \ref{fig:LC_model_comp} presents the posterior parameter distributions of the four \emph{GP-transit} models. As expected, the period and ephemeris are essentially insensitive to the choice of model and underlying assumptions.
As previously mentioned, the limb darkening coefficients of models A and B are theoretically constrained given the assumed stellar temperature, surface gravity and cluster metallicity, whereas they are unconstrained in models C and D; the differences can be seen in the limb darkening panels of Figure \ref{fig:LC_model_comp}.
In the distributions for the radius sum ($(R_{p}+R_{*})/a$), radius ratio ($R_{p}/R_{*}$), inclination ($i$), and the combination terms $\sqrt{e}\cos\omega$ and $\sqrt{e}\sin\omega$, we see the effect of our assumptions about the eccentricity and stellar density. 
First, the stellar density: this essentially places a prior on the transit duration, which in turn constrains the radius ratio, radius sum and $\sqrt{e}\sin\omega$ (and to lesser extents $\sqrt{e}\cos\omega$ and the orbital period, $P$). Most importantly, model A (black), which places a tight constraint on the stellar density, constrains the transit duration such that the model favors a single-peaked radius ratio distribution around $R_{p}/R_{*}\sim0.075-0.08$. Relaxing the constraint on the stellar density (models B and C, orange and blue, respectively) allows the model to explore an additional family of solutions comprising a larger planet-to-star radius ratio (and correspondingly these models explore a larger range of inclinations, radius sums and eccentricities to remain consistent with the observed fluxes). 
The effect of eccentricity can be investigated by comparing models C and D (blue and green, respectively). Forcing the orbit to be circular has a similar effect to placing a tight constraint on the stellar density: model D favors a single-peaked radius ratio distribution and has tighter posterior distributions for the radius sum and inclination than models B and C. 

The GP hyperparameters are mainly insensitive to the chosen model. For the \ktwo\ light curve, the amplitude ($A_{K2}$), period ($P_{\rm{ESS ~ K2}}$) and white noise term ($\sigma_{\rm{K2}}$) are well-constrained by the data but the scale factor ($\Gamma_{\rm{ESS ~ K2}}$) and evolutionary timescale ($l_{\rm{SE ~ K2}}$) have Gaussian priors that essentially define their posterior distributions.
The evolutionary timescale is related to the half-life of the active regions that are responsible for the observed modulation pattern; however, the exact nature of this relationship is not firmly established. It is worth noting that data covering two evolutionary periods is needed to robustly constrain this parameter, so it is not surprising that it is not well constrained in this case. However, while the evolutionary timescale and scale factor affect the other GP hyperparameters, they do not affect the final transit parameters significantly.
For the USNO light curve, the GP amplitude ($A_{\rm{USNO}}$) and timescale ($l_{\rm{USNO}}$) hyperparameters are not well-determined; the differences in the timescale posteriors arise from the tightness of the different transit model fits, and are therefore driven by the stellar density and eccentricity constraints.

\subsubsubsection{Investigating the residual variations in the \ktwo\ and USNO light curves}
\label{sec:resid_var}

\emph{K2}: The model presented above for the \ktwo\ light curve seeks to explain the large-scale structure. While it is an acceptable fit to the data, zooming in on small sections of the light curve reveals small-amplitude, short-timescale variations, which could arise from low-level stellar variations or residual systematics. Arguably, adding another GP into the model might account for these but there would be a lot of flexibility in such a composite GP model, and it is not clear that the residual variations are significant throughout the majority of the light curve. Therefore, to test the significance of these residual variations, we first detrended the light curve with respect to the smoothly varying starspot modulations, and then modelled the residual light curve using a Matern-3/2 kernel in our GP-transit model (as previously discussed, this kernel displays a relatively rough behavior, which is suited to the observed residual systematics). It appears as though the information content of the residual light curve is not sufficient to support the additional treatment of the Matern-3/2 kernel. The GP model favors a very small amplitude and either a long timescale over which to vary, or a very short one, depending on the initial guess for the input scale. Neither of these are suited to the observed residuals, which have timescales of a few hours, and which are at times clearly observed above the noise. We therefore did not use a more sophisticated GP kernel for the \ktwo\ light curve.

\emph{USNO}: The residuals of the USNO model are clearly seen in the bottom panel of Figure \ref{fig:LC_ecc_tlg}. The mean GP model smoothly varies over a relatively long timescale,
in contrast to parts of the light curve that show variations over $\lesssim1$ hour timescales. To assess the significance of these variations we re-ran our model and used the GP itself to estimate the full observational uncertainties (rather than simply allowing it to inflate the uncertainties if required). This second model converges on the same variability and transit parameter values as presented above, which suggests that the information content of the light curve is not sufficient to warrant smaller uncertainties; accordingly, we did not investigate the observed variations further. 

It is worth noting, however, that the validity of the observational uncertainties, and hence the analysis above, are subject to the choice of model. Here, the GP model attempts to find the simplest way of explaining the data, \emph{given our choice of covariance kernel}, and in doing so may opt to increase the uncertainties on each data point rather than the complexity of the chosen family of models if the dataset \emph{as a whole} is unable to support the additional complexity. Accordingly, in model A (presented in \S \ref{sec:GP-transit}), the GP opted to slightly inflate the uncertainties by factors of $\sim$1.37 and 1.17 for the \ktwo\ and USNO light curves, respectively (see Table \ref{tab:lc_model_tab}).

\subsubsection{Sanity check: Independent \ktwo\ light curve analysis}
\label{sec:hannu}

We carry out a second independent 
light curve analysis to ensure the robustness of our primary analysis. This analysis uses the \textsc{K2SC}-detrended PDC-MAP light curve and simplifies the approach by assuming white noise. The transits are modelled using \textsc{PyTransit} \citep{Parviainen2015pt} with a quadratic Mandel-Agol model, and \textsc{emcee} is used for posterior sampling, as in the main analysis (\S \ref{sec:GP-transit}).  The files associated with this analysis can be found from GitHub.\footnote{\url{https://github.com/hpparvi/prae1b}}

This analysis consider four cases for orbital eccentricity, stellar density, and $\log g$ (assuming a stellar radius of 0.43~R$_\odot$), listed in Table~\ref{table:sanity_check_priors}. The sampling space is ten-dimensional, comprised of the zero epoch, orbital period, planet-star area ratio, impact parameter, stellar density, $\sqrt e \cos\omega$, $\sqrt e \sin\omega$, two quadratic limb darkening parameters, and average white noise level. All the parameters except the eccentricity and $\log g$ have uninformative priors, and the limb darkening uses the parametrization described by \citet{Kipping:2013}.

\begin{deluxetable}{lccc}
\tablewidth{\textwidth}
\tabletypesize{\footnotesize}
\tablecolumns{4}
\tablecaption{Priors on System Parameters for Analysis in \S \ref{sec:hannu}
\label{table:sanity_check_priors}}
\tablehead{
    \colhead{Case} & \colhead{Eccentricity} & \colhead{Stellar density (g~cm$^{-3}$)} & \colhead{$\log g$}
    }
\startdata
A (constrained $\log g$ 2) & $\mathcal{U}(0, 0.9)$ & $\mathcal{U}(0.5, 15)$ & $\mathcal{N}(\mu=4.81, \sigma=0.08)$\\
B (constrained $\log g$ 1) & $\mathcal{U}(0, 0.9)$ & $\mathcal{U}(0.5, 15)$ & $\mathcal{N}(\mu=4.5, \sigma=0.5)$\\
C (eccentric, unconstrained) & $\mathcal{U}(0, 0.9)$ & $\mathcal{U}(0.5, 15)$ & $\mathcal{U}(3, 8)$\\
D (circular) & $\delta$ & $\mathcal{U}(0.5, 15)$ & $\mathcal{U}(3, 8)$\\
\enddata
\begin{flushleft}
\footnotesize $\mathcal{U}(a, b)$ - uniform density with a minimum $a$ and maximum $b$ \\
\footnotesize $\mathcal{N}(\mu, \sigma)$ - normal density with mean $\mu$ and standard deviation $\sigma$ \\
\footnotesize $\delta$ - Dirac's delta function.
\end{flushleft}
\end{deluxetable}

The posterior sampling starts by creating a parameter vector population of 100 walkers distributed uniformly inside the prior boundaries. We clump the population close to the global posterior maximum using the Differential Evolution global optimization algorithm implemented in \textsc{PyDE}\footnote{https://github.com/hpparvi/PyDE} \citep{Parviainen2016DE}, and initialize the MCMC sampler with the clumped parameter vector population. The sampler is run over 15000 iterations, and then samples from every 100th iteration, after a burn-in set of 2000 iterations is selected to represent the posterior distribution. The thinning factor of 100 iterations was chosen by inspecting the autocorrelation lengths for individual parameters. 

\begin{figure*}[ht]
    \centering
    \includegraphics[width=\textwidth]{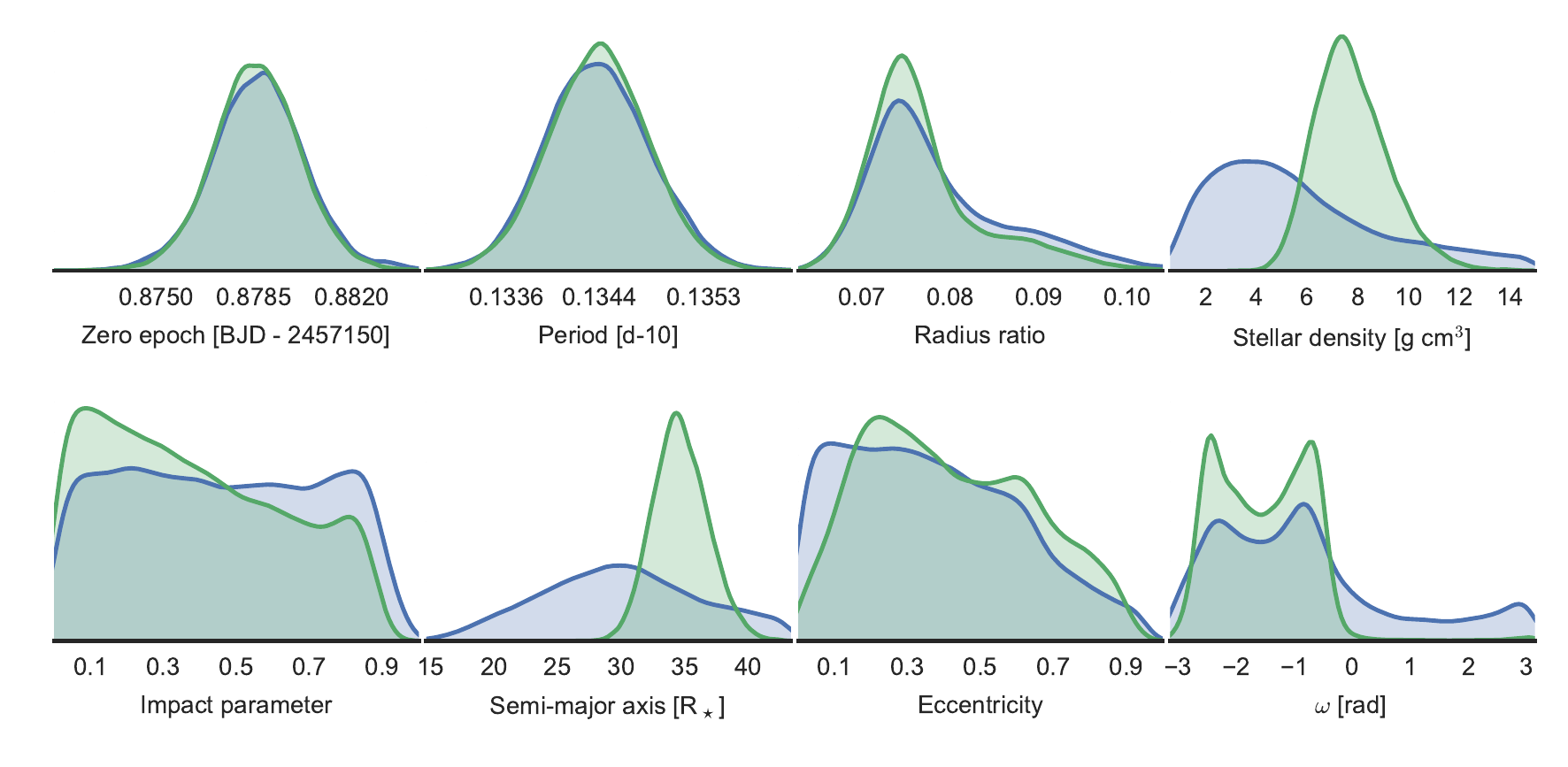}
    \caption{Parameter posteriors from the second \ktwo\ light curve analysis cases A (green) and B (blue), showing the impact of different priors on $\log g$ on the parameter estimates.}
    \label{fig:hannu_pe}
\end{figure*}

The results from the sanity check analyses are shown in Fig.~\ref{fig:hannu_pe}, and they agree with the primary analysis.  One small difference between the two analyses is that the main analysis  finds a broader and near-bimodal distribution for the planet/star radius ratio.  We believe that is because the main analysis uses GPs, whereas the the analysis in this section assumes white noise.  When using GPs, the transit shape does not constrain the parameter space as strongly, and that additional freedom allows for a larger variety of possible solutions.  In the end, we do not believe that this difference represents a challenge to our final system parameters.

\subsection{SED Analysis\label{sec:SED}}

We fit a grid of 
BT-SETTL stellar atmosphere models \citep{Allard:2012} to the available $ugrizJHK_S$ + WISE 1--3 catalog photometry, spanning the wavelength range 0.35--12 $\mu$m (summarized in Table~\ref{tab:phot}).  The reported uncertainties on the SDSS $griz$ photometry are very small ($\sim$0.005~mag), so we rounded up these photometric uncertainties to 0.05~mag to account for photometric variability due to modest activity in this source
as well as zero-point uncertainties of $\approx$0.02~mag in the SDSS photometry.
The free parameters of the fit are $T_{\rm eff}$, $\log g$, [Fe/H], and $A_V$; for the latter we adopted a maximum possible value of 0.085 mag based on the full line-of-sight extinction from the \citet{Schlegel:1998} dust maps. 
We interpolate the model grid by 10~K \teff, from the native gridding of 100~K. We do not interpolate more finely than the native gridding of 0.5 dex in $\log g$ and [Fe/H] because broadband SED fitting is much less sensitive to these parameters compared with \teff.

\begin{deluxetable}{crrl}
\tablewidth{0pt}
\tabletypesize{\footnotesize}
\tablecolumns{4}
\tablecaption{Catalog photometry of \pname \label{table:phot}}
\tablehead{
    \colhead{Bandpass} & \colhead{Mag.} & \colhead{Mag. Err.} & \colhead{Source} 
    }
\startdata
$u$  & 20.704  & 0.053 & SDSS9 \\
$g$  & 18.025  & 0.006 & SDSS9  \\
$r$  & 16.635  & 0.006 & SDSS9 \\
$i$  & 15.370  & 0.005 & SDSS9 \\
$z$  & 14.696  & 0.005 & SDSS9 \\
$J$  & 13.312  & 0.021 & 2MASS \\
$H$  & 12.738  & 0.024 & 2MASS \\
$K$  & 12.474  & 0.021 & 2MASS \\
$W1$ & 12.323  & 0.024 & ALLWISE \\
$W2$ & 12.211  & 0.030 & ALLWISE \\
$W3$ & 11.240  & 0.356 & ALLWISE \\
\enddata
\label{tab:phot}
\end{deluxetable}

\begin{figure}[ht]
    \centering
    \includegraphics[width=\linewidth,trim=75 75 75 75 ,clip]{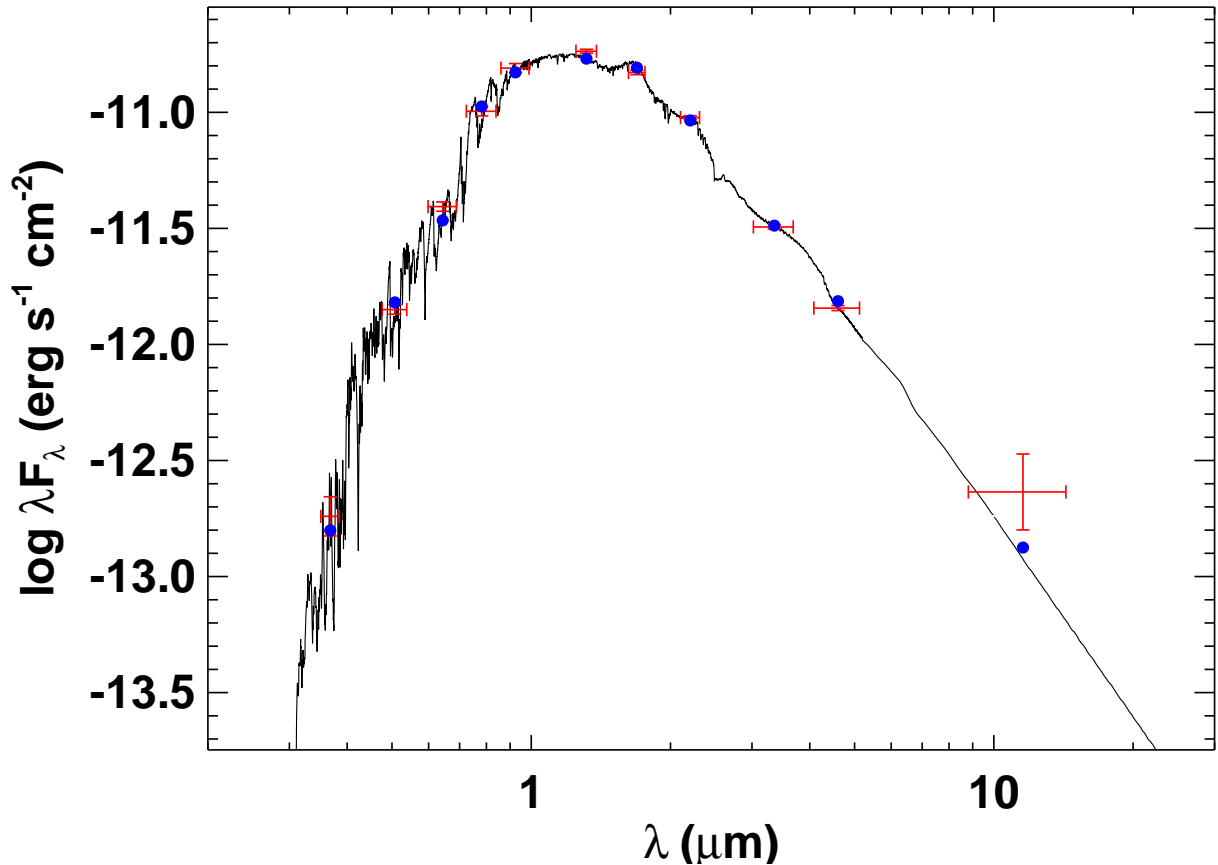}
    \caption{SED fit with BT-SETTL atmosphere model. Red symbols are the catalog photometry, blue symbols are the corresponding model fluxes.}
    \label{fig:sed}
\end{figure}

The resulting best-fit parameters are: 
$T_{\rm eff} = 3350 \pm 50$ K,
$A_V = 0.000^{+0.085}_{-0.000}$,
$\log g = 4.5 \pm 0.5$,
$[Fe/H] = 0.0 \pm 0.5$.
The fit is shown in Figure~\ref{fig:sed}, with a reduced $\chi^2$ of 6.1. 
The uncertainties in the fit parameters are estimated according to the usual criterion of $\Delta\chi^2 = 4.72$, relative to the best fit, for four fit parameters \citep[e.g.,][]{Press:1992}, where we first rescaled the $\chi^2$ so as to make the reduced $\chi^2$ of the best fit equal to 1. This is equivalent to inflating the measurement errors by a constant factor, and has the effect of increasing the parameter uncertainties accordingly.
The $T_{\rm eff}$ from our SED fit is consistent with the previously reported M3.5 spectral type \citep{Adams:2002} to within $\sim$0.5 spectral subtype. 

We can use the fitted SED models to compute a bolometric flux at Earth, again according to the $\Delta\chi^2$ criterion above. This gives $F_{\rm bol}=$ 
2.10$\pm$0.09 $\times 10^{-11}$ erg s$^{-1}$ cm$^{-2}$.
Note that this $F_{\rm bol}$ and its uncertainty are robust because the model fit is essentially an interpolation of the observed fluxes that span the majority of the SED.
This $F_{\rm bol}$ together with
the nominal cluster distance of 182$\pm$6 pc \citep{vanLeeuwen:2009} and the best-fit $T_{\rm eff}$ then permits a calculation of the stellar radius via the Stefan-Boltzmann relation, giving 
$R = 0.44 \pm 0.03$ R$_\odot$. The radius uncertainty includes the uncertainties on the bolometric flux, the parallax distance, and $T_{\rm eff}$, in quadrature. 

This is somewhat larger than the radius predicted from the empirical $T_{\rm eff}$--$R$ relation of \citet{Mann:2016}, which gives 
$R = 0.31 \pm 0.04$
R$_\odot$. However, the Praesepe cluster has been reported to be metal rich, with [Fe/H] $\approx 0.1$ but in some cases reported to be as high as [Fe/H] = 0.27$\pm$0.10 \citep{Pace:2008}. Using this range of [Fe/H] with the $T_{\rm eff}$--[Fe/H]--$R$ relation of \citet{Mann:2016} predicts 
$R = 0.37 \pm 0.03$
R$_\odot$. Finally, instead adopting the  ${M_K}_S$--[Fe/H]--$R$ relation of \citet{Mann:2016}, which those authors find has the tightest of all the empirical relations, gives 
$R = 0.42 \pm 0.01$ R$_\odot$, in good agreement with our measured value. 

Finally, the star is reported both here and in \citet{Barrado:1998} (\S \ref{sec:activity}) to have a filled-in H$\alpha$ line, with EW(H$\alpha$) $\approx$ 0 \AA\, indicating the presence of low-level chromospheric activity. \citet{Stassun:2012} found that the H$\alpha$ EW is directly related to the degree of radius inflation in chromospherically active low-mass stars, and their empirical relations give a radius inflation of $\sim$6\% for EW(H$\alpha$) = 0 \AA. Adjusting the \citet{Mann:2016} ${M_K}_S$--[Fe/H] predicted radius for this amount of inflation, and including the uncertainty in the \citet{Stassun:2012} relation, then gives a final predicted radius of $R = 0.44 \pm 0.02$ R$_\odot$, in excellent agreement with the bolometric radius value we measure above.

\subsection{Planet Properties}
\label{sec:properties}

Next we seek to determine the physical properties of the planet. The planet radius is tied to the stellar radius via the transit radius ratio, and the planet mass is tied to the stellar mass via the radial-velocity mass ratio. The host star's radius we have determined above (Sec.~\ref{sec:SED}) to be $R = 0.44 \pm 0.03$ R$_\odot$, consistent with the (activity-corrected) predicted radius from the empirical relations of \citet{Mann:2016}. Similarly, we estimate the host star mass from the same empirical relations. Using the \citet{Mann:2016} relation for $M_\star$ versus ${M_K}_S$, we obtain an estimated mass of 0.44$\pm$0.04 M$_\odot$, where the uncertainty includes both the photometric and distance uncertainties on ${M_K}_S$ and the scatter in the empirical relation. This mass together with the radius from above then gives a stellar surface gravity of $\log g = 4.82 \pm 0.06$. 

The ratio of the planet radius to the stellar radius is found to be 0.08, and with the measured stellar radius of 0.44 R$_\odot$, that yields a planet radius of 
$0.32\pm0.02$ \rj.  We can also calculate an estimated equilibrium temperature for the planet based on the ratio of the stellar radius to planet semimajor axis, stellar temperature, taking the orbit to be circular, and assuming a Bond albedo of 0.3 (similar to Neptune).  With those assumptions, we find $T_{\rm eq} = 480 \pm 23$ K.  The full list of derived and calculated planet properties is listed in Table \ref{tab:final_properties}.

\begin{figure}
    \centering
    \includegraphics[width=0.8\linewidth]{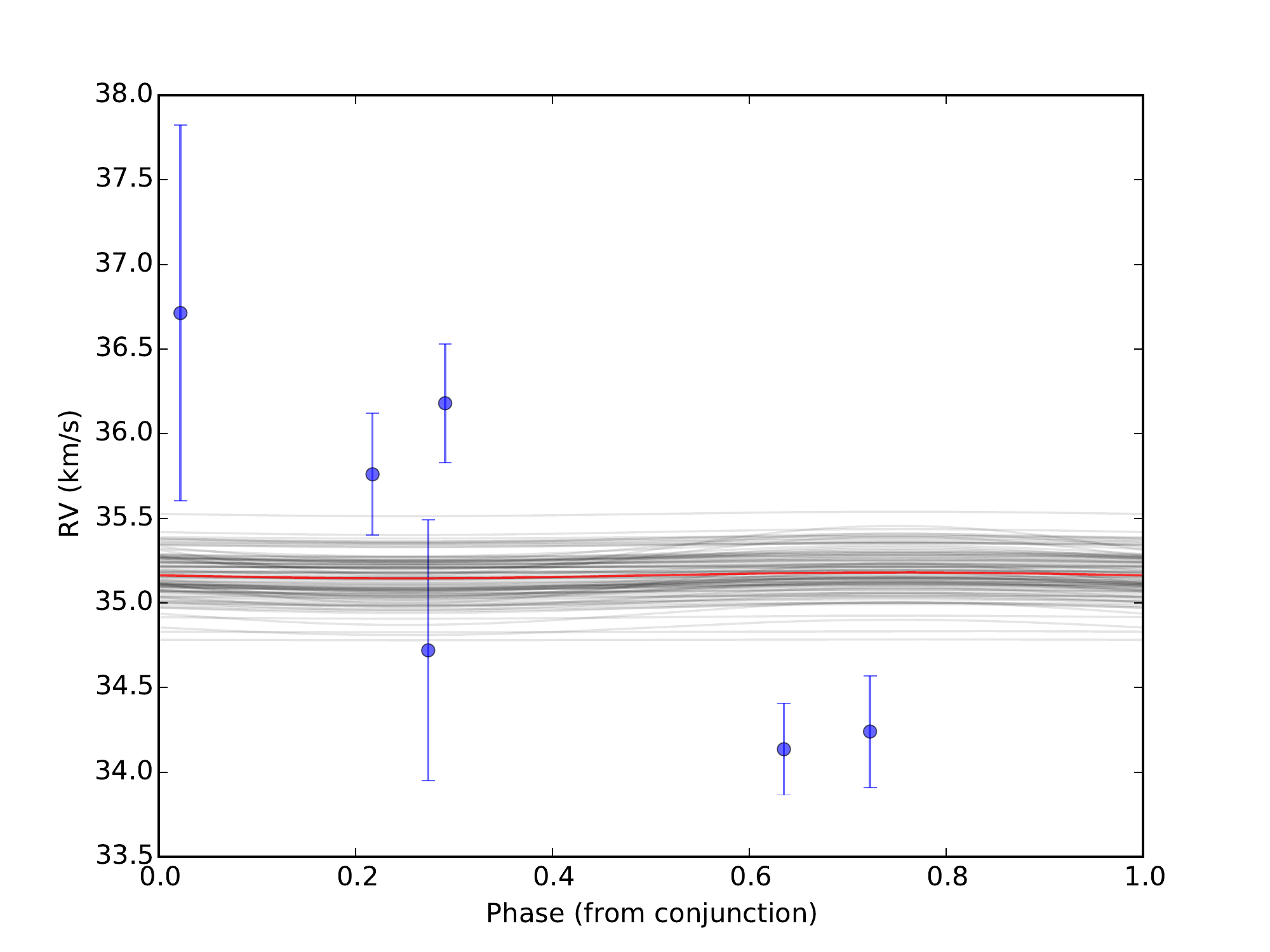}
    \caption{ RV observations of \pname\ in blue, with various fits as described in the text superimposed.  Grey lines represent the individual model fits, with the red line representing the model with parameters equal to the median of the marginalized posterior probabilities.  With large errors compared to the expected RV amplitude, we do not detect the associated dynamical signal of the planet.  Note that the transit observations predict negative RV variation at phase of 0.25, and positive RV variation at phase of 0.75.  We are able to place a limit of 0.83 $M_{\rm Jup}$ at 99.7\% confidence as the upper limit for any planet on the system, still well above the expected mass of the transiting planet detected by \ktwo.}
    \label{fig:maxmass}
\end{figure}

We have performed an analysis of our radial-velocity measurements in order to estimate an upper limit for the mass of the transiting planet.  We have six RV measurements, and the standard deviation of these measurements is 1.08 km\ s$^{-1}$, which corresponds to a maximum mass for the planet of 6.7 $M_{\rm Jup}$ (assuming a zero eccentricity). However, in order to take the shape of the radial velocity model into account, we have tested a circular-orbit model with seven free parameters, namely the systemic radial velocity (uniform prior between 30-40 km s$^{-1}$), orbital period, time of mid-transit and inclination (all three with Gaussian priors based on the light curve analysis), stellar mass (Gaussian prior according to the above estimation), and planet mass (with a uniform prior between 0 and 30 Jupiter masses). We have also included radial velocity jitter (with uniform prior between 0 and 5 km s$^{-1}$) to account for possible additional sources of white noise. We used the \textsc{emcee} code \citep{Foreman-Mackey:2013} to explore the posterior distribution of the planetary mass based on the current data. We used 50 walkers and 5000 steps per walker. We then used a simple burn-in of half of the chains and merge all of them to get the final posterior distribution.  This method is similar to that used in \citet{Grunblatt:2016}.  The result of this analysis yields no detection of the planet (as expected from the radial velocity uncertainties). However, it allows us to set an upper limit for the mass of the transiting object of 0.83 $M_{\rm Jup}$ at 99.7\% confidence given the current data (assuming $e=0$).  The results of these tests are shown in Figure \ref{fig:maxmass}.  Note that in this RV phase plot, the maximum expected RV displacement based on the transit model would occur at a phase of 0.75, opposite the offset of the RV observation at that phase, which therefore strongly constrains the allowable mass of the companion.

The planet orbiting \pname\ can therefore not yet be confirmed through radial velocity detection of the stellar reflex motion.  For a planet with a radius of about 0.32 \rj, we would expect a mass of roughly 15 $R_{\Earth}$ (see \citet[e.g.][]{Chen:2016}).  For a host star with $M_{*} = 0.44\pm0.04$\msun, and a circular orbit with a period of 10.134588 days, we would expect an RV semiamplitude of about 0.008 km s$^{-1}$, far below the $\approx0.3$ km s$^{-1}$ errors of our HIRES RV observations (Table \ref{tab:hires}). 
Although the object would be detectable using current 
state-of-the-art precision radial velocity techniques having  
1-2 m s$^{-1}$ errors \citep[see e.g.][]{Plavchan:2015}, 
the star is too faint in practice for these methods to be applied. 
In such cases, validation of the planet interpretation
is typically undertaken either empirically or statistically.
Empirical methods use a suite of observations that may be sensitive 
to detection of certain kinds of blend scenarios 
\citep[e.g.][]{ODonovan:2006} that could mimic the observed transit signal. 
Statistical methods can estimate the probability of various contaminating
eclipsing binary scenarios given the source location on the sky 
relative to the modelled galactic field star population,
and assumptions regarding the mass function, binary fraction,
and binary mass ratios \citep[e.g.][] {Morton:2014}.

As we were writing this paper, we became aware of work by \citet{Obermeier:2016} and \citet{Mann:2017praesepe},
who independently discovered and followed up 
the same \ktwo\ time series data on EPIC 211916756.  The thorough
false-positive analysis presented in the \citet{Obermeier:2016} paper need not be
repeated here.  From empirical considerations, no companions 
are detected in high spatial resolution imaging or in spectroscopy.
From statistical considerations, a false positive probability 
calculation rules out line-of-sight blended and bound hierarchical
eclipsing binary systems.  Remaining for consideration is the
planet hypothesis.

\begin{deluxetable}{lcc}
\tablewidth{0pt}
\tabletypesize{\footnotesize}
\tablecolumns{3}
\tablecaption{Final properties of the \pname\ system \label{table:final}}
\tablehead{
    \colhead{Property} & \colhead{Value} & \colhead{Uncertainty} 
    }
\startdata
$M_\star$ & 0.44 M$_\odot$ & 0.04 M$_\odot$ \\
$R_\star$ & 0.44 R$_\odot$ & 0.03 R$_\odot$ \\
$T_{\rm eff}$  & 3350 K & 50 K \\ 
$\log g_\star$  & 4.82 & 0.06 \\ 
${\rm [Fe/H]}$  & 0.1 & 0.1 \\
$R_{\rm pl}$  & 0.32 \rj & 0.02 \rj \\
$M_{\rm pl}$  & $<$1.67 \mj & -- \\
$T_{\rm eq}$  & 480 K & 23 K \\
Age &  \multicolumn{2}{c}{$\sim$600$^{a}$ Myr -- {\bf 790$\pm$60}$^{b}$ Myr}  \\
\enddata
\footnotesize \hspace{1in} $^{a}$\citet{Adams:2002}
\footnotesize \hspace{0.5in} $^{b}$\citet{Brandt:2015}
\label{tab:final_properties}
\end{deluxetable}

\section{Discussion\label{sec:disc}} 

\begin{figure}
    \centering
    \includegraphics[width=0.8\linewidth]{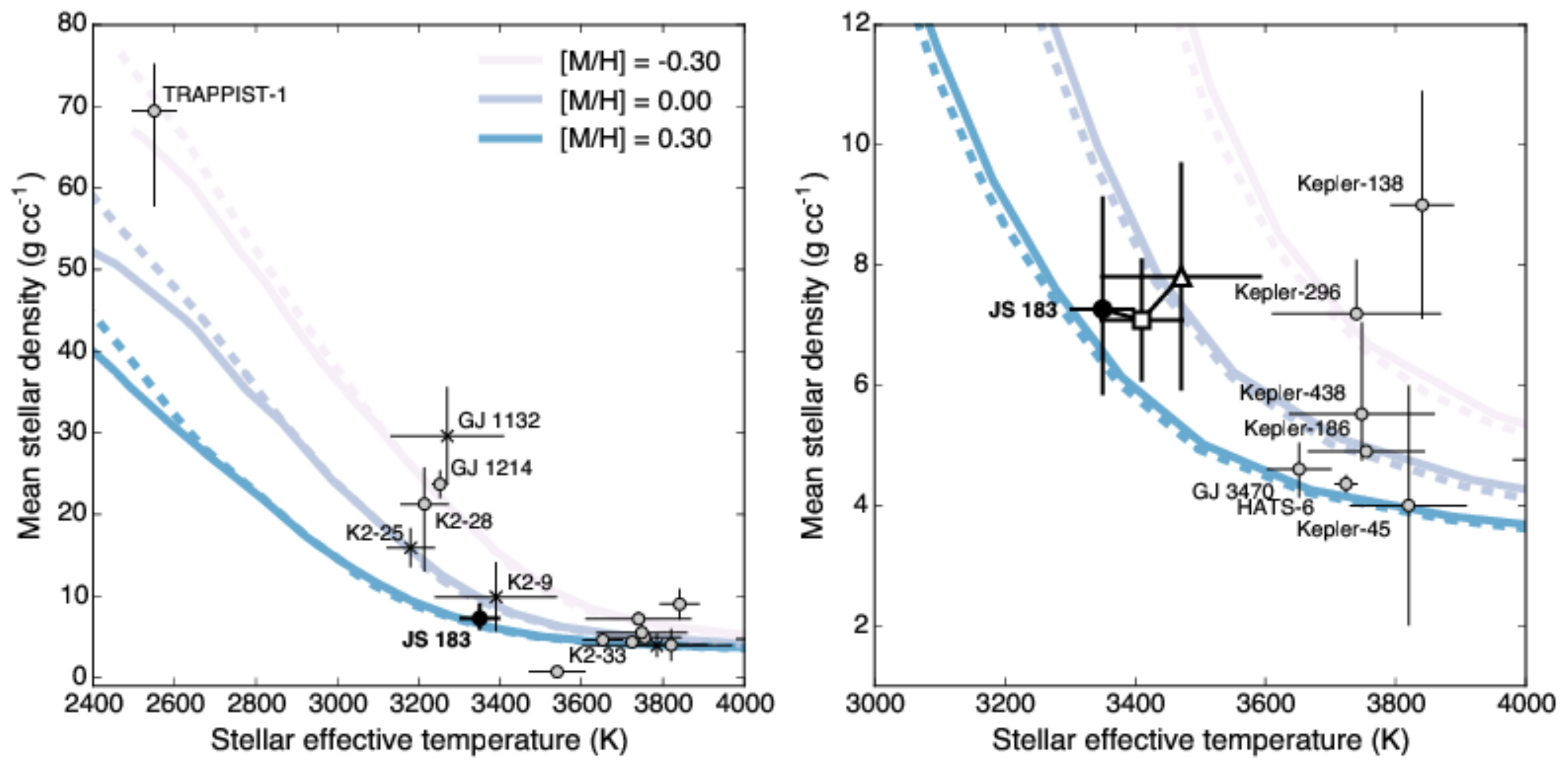}
    \caption{Left: Late-type exoplanet host stars with well-determined stellar densities from transit fitting (filled markers) or through some combination of spectroscopy, models, and empirical relations (x's). Underplotted are PARSEC v1.2S isochrones \citep{Bressan:2012} in the effective temperature--mean stellar density plane. Solid and dashed curves represent 600 Myr and 1 Gyr isochrones, respectively. Curve colors correspond to different metallicities. The solar metallicity isochrone reproduces well the majority of transiting planet host stars cooler than 4000 K. At these cool temperatures, the mean stellar density is largely an indicator of metallicity and is relatively insensitive to age. The parameters for GJ 1214 are adopted from \citet{AngladaEscude:2013}. Preliminary transit fitting results of GJ 1132 suggest a lower stellar density than the value from empirical relations, which is plotted here (Berta-Thompson, private comm.). The planet host K2-33 is a pre-main sequence star \citep{David:2016b, Mann:2016b} and thus substantially less dense than the field population. Right: an enhanced view of the figure at left in the region around JS 183, only with those exoplanet hosts with densities constrained by transit fits. Three determinations for the parameters of JS 183 are indicated by the filled black circle (this work), open square \citep{Mann:2017praesepe}, and open triangle \citep{Obermeier:2016}. Notably, a super-solar metallicity isochrone best matches the parameters of JS 183, consistent with the enhanced metallicity of the Praesepe cluster.  Exoplanet host star data were culled from the NASA Exoplanet Archive, but mean stellar densities were taken from the transit fitting results of the primary references listed in the Exoplanet Archive, wherever reported.}
    \label{fig:rad_metal}
\end{figure}

For low-mass stars, metallicity substantially affects the radius of the star \citep[e.g.][]{Mann:2016}.  Since the planet radius as obtained from the transit depends directly on the stellar radius, its precision is limited by the precision of the stellar radius, and hence metallicity. In general, cluster stars such as \pname\ have better constrained metallicities than field stars, given the ability to study a large population that is presumed to be chemically homogeneous. 

To date, there are more than 2500 unique hosts to confirmed exoplanets, but only 100 with effective temperatures $\leq$4000 K. Of these, only 16 have reliable mean stellar densities determined either from transit fitting or detailed stellar characterization and reported in the NASA Exoplanet Archive\footnote{\url{http://exoplanetarchive.ipac.caltech.edu/}}. We examine this complete sample of late-type hosts to confirmed transiting exoplanets in Figure \ref{fig:rad_metal}. The figure shows that the higher metallicity of \pname\ leads to a larger radius and thus smaller stellar density than other transiting planet hosts of comparable temperature. For low-mass stars, which evolve slowly relative to higher masses, the mean stellar density is largely an indicator of metallicity and is relatively insensitive to age. Typically, as in this work, a stellar density determined from some combination of empirical relations and models is used as a prior in exoplanet transit fitting, due to the covariances between stellar density, $e$, and $\omega$. However, if the orbital elements of an exoplanet are well constrained, the stellar density inferred from a transit light curve can be a powerful diagnostic of metallicity and test of stellar models at low masses; the technique of using exoplanet transits to determine precise stellar densities is sometimes called asterodensity profiling~\citep{Kipping:2014}.

Late-type transiting exoplanet hosts with temperatures higher than 3400~K are in general agreement with a field age, solar-metallicity isochrone, using the PARSEC v1.2s models~\citep{Bressan:2012}. However, the majority of transiting exoplanet hosts with temperatures below 3400~K possess transit-derived stellar densities that are too large to be reproduced by the solar metallicity models. While transit-derived stellar densities are often poorly constrained due to the unknown orbital elements, transit fits tend to drive stellar density to \textit{lower} values due to the aformentioned covariances and a bias favoring high impact parameters or grazing transits~\citep{Kipping:2016}. There are hints of this bias apparent in Figure~\ref{fig:hannu_pe}. Thus, $e$ and $\omega$ may not be the only culprits explaining the larger dispersion in stellar densities for the coolest exoplanet hosts. It is possible that 1) stellar models may underpredict the densities of the coolest stars, or 2) some of the the cool transiting exoplanet hosts have significantly sub-solar metallicities. We presume the latter explanation is unlikely, especially given the fact that the well-studied planet host GJ 1214 has a \textit{super-solar} metallicity~\citep{RojasAyala:2012}.

In comparison with other well-studied exoplanet hosts with a similar effective temperature, JS 183 has the lowest mean stellar density to date\footnote{K2-9 has a similar temperature and bulk density, though \citet{Schlieder:2016} note a discrepancy between the transit-derived stellar density and the spectroscopically determined value which we show in the figure.}. The star's low density is consistent with expectations of an inflated radius due to the Praesepe cluster's super-solar metallicity. 
The fact that JS~183's spectral type of M3.5 right near the stellar fully convective boundary provides the intriguing suggestion that the models are performing well at masses/temperatures just above that boundary but performing poorly below that boundary, where the onset of full convection leads to additional challenges for low-mass stellar modeling.

The recent discoveries of short-period ($<10$s days) Neptune-sized planets around young low-mass stars reveal planet radii significantly larger than the planets observed around field M dwarfs \citep{Dressing:2013,Dressing:2015}, even after accounting for the trend of planet inflation with insolation. One possibility is that the young planets have not finished contracting, and will eventually settle into the population of short-period super-Earths, a region of higher occurrence, as determined from Kepler statistics of field stars.

\section{Summary and Conclusions\label{sec:summary}} 

We have reported here the discovery of a Neptune-size planet transiting an M dwarf member of the Praesepe open cluster, also discussed by \citep{Obermeier:2016} and \citet{Mann:2017praesepe}.  Although we do not have a dynamical mass of the companion, we can constrain its mass to be planetary.  This discovery adds to the small list of exoplanets found in open clusters, and planetary companions of M dwarfs.  The well-known age and metallicity of the Praesepe cluster give reliable values for those properties of the planet, which can be valuable for consideration of theoretical models of planet formation and evolution. 

This system also features the best-determined radius for an M dwarf planet host, and is only the third M dwarf planet host for which a reliable age is known via cluster membership.
Indeed, with the lowest stellar density known among transiting planet hosts near the stellar fully convective boundary, JS~183 suggests that current stellar models are able to reproduce well its properties as arising from its super-solar metallicity, and further suggests that models remain challenged for stars with masses below the fully convective boundary.

\acknowledgements
J.P. would like to thank Ian Crossfield for useful discussions.  
T.J.D. is supported by an NSF Graduate Research Fellowship under Grant DGE1144469.   B.J.S.P. would like to thank Balliol College and the Clarendon Fund for their financial support of this work.  S.A. and H.P. acknowledge funding from the Leverhulme Trust. S.A. received support from the UK Science and Technology Facilities Council (STFC).
K.G.S. acknowledges partial support through NSF PAARE grant AST-1358862.
J.L-B. acknowledges support from the Marie Curie Actions of the European Commission (FP7-COFUND).

\bibliography{ms}

\end{document}